\documentclass[epj]{svjour}
\usepackage{latexsym}
\usepackage{graphics}
\RequirePackage{graphicx}
\RequirePackage{fix-cm}
\usepackage{color}
\begin{document}
\title{Dimesonic states with the heavy-light flavour mesons}
\subtitle{}
\author{D. P. Rathaud \and Ajay Kumar Rai 
\thanks{\emph{Present address:} dharmeshphy@gmail.com; raiajayk@gmail.com}%
}                     
\offprints{}          
\institute{Department of Applied Physics, Sardar Vallabhbhai National Institute of Technology, Surat - 395007, Gujarat, INDIA 
}
\date{Received: date / Revised version: date}
%
\abstract{
In this work, we have calculated the mass spectra and decay properties of dimesonic states in the variational scheme. The inter-mesonic interaction considered as the Hellmann potential and One Pion Exchange potential. The mass spectra of the $D\overline{D^{*}}$, $D^{*}\overline{D^{*}}$, $D\overline{B^{*}}$, $B^{*}\overline{D}$, $B\overline{B^{*}}$, $B^{*}\overline{B^{*}}$ bound states are calculated. The states X(3872), $X_{2c}(4013)$, $Z_{b}(10610)/X_{b}$ and $Z_{b}(10650)/X_{b2}$ are compared with $D\overline{D^{*}}$, $D^{*}\overline{D^{*}}$, $B\overline{B^{*}}$  and  $B^{*}\overline{B^{*}}$ dimesonic bound states. {\bf To probe the molecular structure of the compared states, we have calculated the decay properties sensitive to their long and short distance structure of the hadronic molecule. The radiative decay for the state X(3872) into $J/\psi \gamma$ and $\psi(2S) \gamma$ have been calculated and the ratio is found to be ten times lesser than the experimental value whereas the other decay modes are comparable with other theoretical and experimental results. This results restrict us to assigned the pure molecular structure to the X(3872). But, Our results suggests that the compared states are close to the  molecular structure or have dominant molecular component in their wave function. Apart from these, the other calculated mass spectra of dimesonic states are predicted and for such bound states, the experimental search are suggested.}\\
} 
\maketitle
\section{Introduction}
\label{intro}
In 2003, Belle reported the discovery of a charmonium like neutral state X(3872) with mass=$3872\pm0.6\pm0.5$ MeV and width $<2.3$ MeV \cite{Olive,S.K. Choi} which latter confirmed by DO \cite{V.M. Abazov}, CDF \cite{D. Acosta} and BARBAR \cite{B. Aubert}. This discovery fed excitement in the charmonium spectroscopy due to unconventional properties. The state could not be explained through ordinary meson($q\overline{q}$) and baryon (qqq) scheme, yet. Indeed, the conventional theories predicts complicated color neutral structures and search of such exotic structures are as old as quark model \cite{M. Gell-Mann,G. Zweig}. After the discovery of X(3872), the large numbers of charge, neutral and vector states have been detected in various experiments and famous as the XYZ states. Recently, the charge bottomoniumlike resonances $Z_{b}(10610)$ and $Z_{b}(10650)$ have been reported by Belle Collaboration in the process $\Upsilon(5S) \rightarrow \Upsilon(nS) \pi^{+} \pi^{-}$ and $\Upsilon(5S) \rightarrow h_{b}(mP) \pi^{+} \pi^{-}$ \cite{Bonder}. Moreover, a another state reported by BESIII Collaboration \cite{M. Ablikim-1} as a $Z_{c}(4025)$ in the $e^{+}e^{-}\rightarrow h_{c} \pi^{+} \pi^{-}$ reaction. Again, the BESIII Collaboration reported one more state as a $Z_{c}(3900)$ from invariant mass $J/\psi \pi$  in the $e^{+}e^{-}\rightarrow \pi^{-}\pi^{+}J/\psi$ reaction \cite{M. Ablikim-2}, whereas the Belle \cite{Z. Q. Liu} and CLEO \cite{T. Xiao} reconfirmed the status of these states. The sub structure of the all these states are still a open question. They might driven the exotic structure like tetraquark, molecular or hybrid as predicted by theory of the QCD and needs extra theoretical attentions. \\

In the present study, we have focused on the molecular structure (meson-antimeson bound state, just like deuteron). The multiquark structures have been studied since long time \cite{R.L. Jaffe,R.L. Jaffe-2,J. weinstein}. Nils T$\ddot{o}$rnqvist predicted mesonic molecular structures and introduced it as $dusons$ by using one pion exchange potential \cite{Tornqvist,Tornqvist-Z}. With heavy flavour mesons, various authors predicted the bound state of $D\overline{D^{*}}$ and $B\overline{B^{*}}$ as a possible mesonic molecular structures as well as studied the possibilities of the $D^{*}\overline{D^{*}}$ and $B^{*}\overline{B^{*}}$ as Vector-Vector molecule \cite{Tornqvist,Tornqvist-Z,Thomas,Gui-Jun Ding,Mahiko Suzuki,F. E. Close}, also it have been studied in various theoretical approaches like potential model \cite{Tornqvist,Tornqvist-Z,Thomas,Gui-Jun Ding,Rathaud,Rai,Smruti}, effective field theory \cite{S. Fleming,Miguel,Feng-Kun Guo}, local hidden gauge approach \cite{R. Molina}, effective range expansion \cite{Kang} etc..  \\

In the variational scheme, we have used the potential model approach to study the meson-antimeson bound system. For that, we have used the effective potential (Hellmann potential)\cite{H. Hellmann} with One Pion Exchange Potential (OPEP). Here, the Hellmann Potential is the pseudopotential (superposition of the Coulomb plus Yukawa). This effective potential is used as an approximation for the simplified description of the complicated system. The Hellmann potential had used previously to explain polar molecules and two particle interaction bound states \cite{H. Hellmann,Adamowski}.  

We assume that  the color neutral states experience residual force due to confined gluon exchange between quarks of the hadrons, Skyrme-like interaction. As mentioned by Greenberg in Ref. \cite{Greenberg} and also noted by Shimizu in Ref \cite{Shimizu} that this  dispersive force or the attraction between color singlet hadron comes from the virtual excitation of the color octet dipole state of each hadron \cite{Greenberg,Shimizu}. Indeed, long ago Skyrme \cite{Skyrme} in 1959 and then Guichon \cite{Guichon}, in 2004  had remarked that the nucleon internal structure to the nuclear medium does play a crucial role in such strong effective force of the N-N interaction. In the study of the s-wave N-N scattering phase shift, in Ref.\cite{Khadkikar}, Khadkikar and Vijayakumar used the color magnetic part of the  Fermi-Breit unconfined one-gluon-exchange potential responsible for short range repulsion and sigma and pion are used for bulk N-N attraction, the residual interaction have been used for the study of di-hadronic molecular system in \cite{rai2006}. Thus, with such assumption of the interaction, the mass spectra of the dimesonic bound states are calculated in this study. \\

For molecular binding, the Ref.\cite{Swanson} found that the quark exchange alone could not bind the system, led to include one pion exchange. The Ref.\cite{Tornqvist-Z} mention some additional potential strength required with one pion exchange. Whereas, the dynamics at very short distance led to complicated heavy boson exchange models as studied in \cite{Gui-Jun Ding}. In all these studies \cite{Tornqvist-Z,Thomas,Gui-Jun Ding,Swanson}, one common conclusion was extracted that the highly sensitive dependence of the results on the regularization parameter. To avoid these dependency and complicated heavy boson exchange in this phenomenological study, we used the Hellmann potential in accordance to delicate calculation of attraction and repulsion at short distance. The overall Hellmann potential represents the residual strong interaction  at short distance in favor of the virtual excitation of the color octet dipole state of each color neutral states. The  OPEP is included for long range behavior of the strong force. The OPE potential could be split into two parts (i) central term with spin-isospin factor and (ii) tensor part. We have analyzed the effect of these two parts. Whereas, the tensor term to be found play a very crucial role,  implicit the necessity of it. Hence, the bound state of $D\overline{D^{*}}$ is compared with the state X(3872) which also have been predicted as mesonic molecule by the authors of Ref.\cite{Tornqvist,Gui-Jun Ding,Thomas,S. Fleming,Feng-Kun Guo,Miguel} whereas the states $Z_{b}(10610)$ and $Z_{b}(10610)$ which are close to the $B\overline{B^{*}}$ and $B^{*}\overline{B^{*}}$ threshold.\\

To test the internal structure of the state, in general, one have to look for the decay pattern of the state. In Ref. \cite{Miguel,Feng-Kun Guo,Swanson,N.N. Achasov,Miguel-L,F. Aceti,Eric Braaten}, the hadronic decays of the X(3872) have been studied in accordance to its decay mode sensitive to the short or long distance structure of the state. To test the compared states as dimesonic system, we have used the binding energy as input for decay calculation. We have adopted the formula  developed by authors of Ref. \cite{Feng-Kun Guo} for the partial width sensitive to the long distance structure of the state, whereas, the formula for the decay mode sensitive to short distance structure of the state is taken from \cite{Rui Zhang}. In Ref. \cite{Miguel}, authors predicted existence of the neutral spin-2 ($J^{PC}$=$2^{++}$) partner of X(3872), would be $D^{*}\overline{D^{*}}$ bound state, and in same way expected spin-2 partner of $B\overline{B^{*}}$, would be  $B^{*}\overline{B^{*}}$ bound state, on the basis of heavy quark spin symmetry and calculated the hadronic and radiative decays. We have used formula developed by \cite{Miguel} for the radiative decay calculation. With calculated binding energy, the decay properties are in good agreement with \cite{Miguel,Feng-Kun Guo}. \\

The article is organize as follows, after the brief introduction we have presented the theoretical framework for the calculation in the sec-II then in sec-III we present the results for the deuteron form the model and generalize our approach. In sec-IV and V, we presents our calculated mass spectra and the decay widths for dimesonic states and finally we summaries our present work in sec-VI.  
     
\begin{table*}
\caption{Masses of the mesons (in MeV)\cite{Olive}}
\begin{tabular}{cccccccc}
\hline
Meson & $D^{\pm}$ & $D^{0}$ & $D_{s}^{\pm}$ & $B^{\pm}$ & $B^{0}$ & $B_{s}$ & $B_{c}^{\pm}$ \\
Mass & 1869.62 & 1864.86 & 1968.49 & 5279.25 & 5279.58 & 5366.77 & 6277 \\
\hline
Meson & $D^{*0}$ & $D^{*\pm}$ & $D_{s}^{*\pm}$ & $B^{*}$ & $B_{s}^{*}$ \\
Mass & 2006.98 & 2010.28 & 2112.30 & 5325.20 & 5415.40 \\
\hline 
\end{tabular}
\end{table*}

\section{Theoretical Framework}
To extract the s-wave ground state energy of the dimesonic (meson-antimeson) system, we have employed the variational scheme to solve the Schr$\ddot{o}$dinger equation and for that the hydrogenic trial wave function is being used. The Hamiltonian of the system is read \cite{Rathaud,Rai,Devlani}
\begin{equation}
H=\sqrt{P^2+m_{d}^{2}}+\sqrt{P^2+m_{b}^2}+V(r_{12})
\end{equation}
Here, $m_{d}$ and $m_{b}$ are the masses of constituents and P is the relative
momentum of two mesons while the $V(r_{12})$ is the inter mesonic interaction potential
of the meson-antimeson system. In the present study, we incorporated the heavy light mesons (contain the u,d,s,c,b flavour quarks or antiquarks) for dimesonic systems. To incorporate the relativistic effect due to the light quarks we includes correction to the potential as well as expand the kinetic energy term of the Hamiltonian up to $\cal{O}$($P^{6}$). The binomial expansion of the kinetic energy term reads  

\begin{eqnarray}
K.E. &=& \frac{P^{2}}{2}\left(\frac{1}{m_{d}} +{\frac{1}{m_{b}}}\right)
- \frac{P^{4}}{8}\left(\frac{1}{m_{d}^{3}} +{ \frac{1}{m_{b}^{3}}}\right) 
+ \frac{P^{6}}{16}\left(\frac{1}{m_{d}^{5}} +{ \frac{1}{m_{b}^{5}}}\right)+{\cal{O}}(P^{8})
\end{eqnarray}

and the dimesonic interaction potential reads

\begin{equation}
V(r_{12})=V^{0}(r_{12})+\left(\frac{1}{m_{d}}+\frac{1}{m_{b}}\right) V^{1}(r_{12})
\end{equation}
where the term $V^{0}(r_{12})$ and $V^{1}(r_{12})$ are given by 

\begin{eqnarray}
V^{0}(r_{12})&=&V_{h}(r_{12})+V_{\pi}(r_{12})\nonumber  \\ &and& \nonumber \\
V^{1}(r_{12})&=&-\frac{C_{F}C_{A}}{4} \frac{{k_{mol}}^{2}}{r_{12}^{2}}+{\cal O}\left(\frac{1}{m^{2}}\right) 
\end{eqnarray}
Here, $V_{h}{(r_{12})}$ and $V_{\pi}(r_{12})$ are the Hellmann and One Pion Exchange Potential (OPEP) respectively. The $V^{1}(r_{12})$ is the relativistic mass correction where $C_{F}=4/3$ and $C_{A}=3$ are the Casimir charges of the fundamental and adjoint representation respectively \cite{Koma2006}. \\ 
The Hellmann potential takes the form, namely
\begin{equation}
V_{h}{(r_{12})}=-\frac{k_{mol}}{r_{12}}+B\frac{e^{-C r_{12}}}{r_{12}}
\end{equation}
Here, $k_{mol}$ is the residual coupling constant of the coulombic part of the potential whereas B and C are the constant of the Yukawa part of the potential. The determination of the constant B and C for the case of dimesonic states could be extracted by relation between mass of dimesonic state with mass of the deuteron. If $ m_{m}$ = n $m_{dut}$, where $m_{m}$ is the threshold mass of dimesonic state and $m_{dut}$ mass of deuteron and n is integer number then we arrived on relation such that
\begin{eqnarray}
B &=& \frac{B_{0}}{n}  ; \ \ \ \ C = n C_{0}
\end{eqnarray}
Here, $B_{0}$ and $C_{0}$ are the constant to fit empirical value of the binding energy for the deuteron.    
Whereas, the $V_{\pi}(r_{12})$ (OPEP) takes the form \cite{Paris,Pandharipande}, namely 
\begin{equation}
V_{\pi}(r_{12})=V_{\pi}(c)+V_{\pi}(t)
\end{equation}
where the term $V_{\pi}(c)$ and $V_{\pi}(t)$ are given by
\begin{eqnarray}
V_{\pi}(c)&=& \frac{1}{12}\frac{g_{0}^{2}}{4\pi}\left(\frac{m_{\pi}}{m_{m}}\right)^{2}\left(\tau_{1}\cdot\tau_{2}\right)\left(\sigma_{1}\cdot\sigma_{2}\right)\left(\frac{\Lambda^{2}}{\Lambda^{2}-m_{\pi}^{2}}\right) \nonumber  
\\& &  
\left(\frac{e^{-m_{\pi}r_{12}}}{r_{12}}-\left(\frac{\Lambda}{m_{\pi}}\right)^{2}\frac{e^{-\Lambda r_{12}}}{r_{12}}\right)
\end{eqnarray}

\begin{eqnarray}
V_{\pi}(t)&& = \frac{1}{12}\frac{g_{0}^{2}}{4\pi}\left(\frac{m_{\pi}}{m_{m}}\right)^{2}\left(\tau_{1}\cdot\tau_{2}\right)(S_{12})\left(\frac{\Lambda^{2}}{\Lambda^{2}-m_{\pi}^{2}}\right) \nonumber  
\\& & 
\left[\left(1+\frac{3}{m_{\pi}r_{12}}+\frac{3}{m_{\pi}^{2}r_{12}^{2}}\right)\frac{e^{-m_{\pi}r_{12}}}{r_{12}} \right. \nonumber \\
&& \left.-\left(\frac{\Lambda}{m_{\pi}}\right)^{2}\left(1+\frac{3}{\Lambda r_{12}}+\frac{3}{\Lambda^{2}r_{12}^{2}}\right)\frac{e^{-\Lambda r_{12}}}{r_{12}}\right]
\end{eqnarray}

The detail discussion on the  $V_{h}(r_{12})$ and $V_{\pi}(r_{12})$ potentials are discussed with their parameters in the appendix (see Appendix A and B). Here,  $V^{1}(r_{12})$ is the relativistic mass correction included in the potential. This correction originally studied by Y. Koma where the relativistic correction to the QCD static interquark potential at  ${\cal{O}}(1/m)$ was investigated non-perturbatively. The non-perturbative  form of  $V^{1}(r_{12})$ is not yet known\cite{Koma2006}. These correction is found to be similar to the Coulombic term of the static potential when applied to the charmonium and to be one-forth of the Coulombic term for the bottomonium \cite{Koma2006}. Usually, 
the  static  potential obtained by evaluating the expectation value of the Wilson loop where as the leading and next-to-leading order corrections, which are classified in powers of the inverse of heavy-quark mass \cite{Koma2006}. Taichi Kawanai in Ref.\cite{Kawanai} had presented an investigation of the interquark potential determined from the $q\overline{q}$ Bethe-Salpeter (BS) amplitude for heavy quarkonia in lattice QCD. In their approach, the Coulombic parameter in the static potential (Cornell Potential ) depends on the quark mass significantly and observed that there is no appreciable dependence of the quark mass on the string tension. These observations agree with several features of the $1/m_{q}$ corrections to the static $q\overline{q}$ potential found in the work of Koma \cite{Koma2006}. 
In the Ref.\cite{Pedro1,Pedro2}, the authors have studied the tetraquark systems within the Born-Oppenheimer approximation with finite quark mass. In this study they mentioned that it is interesting but difficult to incorporate the correction with  static potential.
Recently, we have used this correction for study of $B_{c}$ meson \cite{Devlani}, baryons \cite{Zalak} and dimesonic states \cite{Rathaud}. To test the effect of the correction in the dimesonic systems we incorporate it in the meson-antimeson interaction potential with the residual coupling constant. The effect of the expansion of the kinetic energy part as well as correction added in the potential are discussed in the section-3. \\

In our approach, we have used the hydrogenic trial wave function read as
\begin{equation}
R_{nl}(r)=\left(\frac{\mu{}^{3}(n-l-1)!}{2n(n+l)^{3}!}\right)^{\frac{1}{2}}\left(\mu r_{12}\right)^{l}e{}^{\frac{-\mu r_{12}}{2}}L_{n-l-1}^{2l+1}(\mu r_{12})
\end{equation}
Here, the $L_{n-l-1}^{2l+1}(\mu r_{12})$ is the Laguerre polynomial and $\mu$ is the variational parameter. In the variational approach, the ground state energy of the low-lying system is calculated by obtaining the expectation value of the Hamiltonian. The variational parameter ($\mu$) is determined for each state by using the Virial theorem \cite{Rathaud,Devlani}
\begin{eqnarray}
H\psi &=& E\psi \nonumber \\  & \nonumber and & \\
\left\langle K.E.\right\rangle &=& \frac{1}{2} \left\langle \frac{r_{12}dV(r_{12})}{dr_{12}} \right\rangle
\end{eqnarray}
The parameters of the $V_{h}(r_{12})$ are determined through the Eq.(6), for all dimesonic calculations. In which, the constant $B_{0}$=6.25 and $C_{0}$=0.235 are fitted for the binding energy of the deuteron and taken as the input for the Eq.(6). For the OPE potential, the parameters are taken as per discussed in the Appendix-B. The masses of the mesons used in the calculations are taken from Particle Data Group \cite{Olive}, 
tabulated in the Table-1.  \\

\subsection{Decay width}
The study of the decay properties is very important to get informations about the  internal structure of the mesonic molecule. In general, it is preferable to probe the decay processes with one of the constituent meson in the final state and rest of the final constituent decay into other particles (or photon). While some decay channels, mainly, sensitive to the detail of the wave function at the short distances, may be smaller than the size of the molecule. We have attempted the calculation for decay properties of (Mainly, $D\overline{D^{*}}$, $D^{*}\overline{D^{*}}$, $B\overline{B^{*}}$, $B^{*}\overline{B^{*}}$. Wherein, in the literature, most of the molecular structures are predicted) dimesonic states. We have adopted the formula form Ref.\cite{Rui Zhang}, for the calculation of the decay channel sensitive to the short distances. Moreover, the formula have been derived by Ref\cite{Feng-Kun Guo} and \cite{Miguel} for the decay channel  sensitive to the long distance part and decaying into mesons or photon, are being used for hadronic and radiative decays respectively.    \\

(1) The partial decay width $\Gamma_{i}$ (sensitive to short distance) is given by \cite{Rui Zhang}\\
$d + b \rightarrow M \rightarrow c+e \rightarrow c+f+h$
\begin{equation}
\Gamma_{i} =  \frac{\ \left|\psi(0)\right|^{2}l}{16\pi m_{m}^{3}} \left|{\cal{M}}\right|^{2}
\end{equation}
Here, $\psi(0)$ is the wave function at the origin, $m_{m}$ is the mass of dimesonic molecule and $l$ is the magnitude of 3-momentum of the decay product and  $\cal{M}$ is the amplitude 
are given by 

\begin{eqnarray}
l^{2} &=& \frac{\lambda(m_{m}^{2},m_{c}^{2},m_{e}^{2})}{4m_{m}^{2}} \nonumber \\
{\cal{M}}&=&\frac{ k_{mol}^{2}}{q^{2}-m_{q}^{2}}\left(\frac{\Lambda_{1}^{2}-m_{q}^{2}}{\Lambda_{1}^{2}-{q}^{2}}\right)
\end{eqnarray}

$\lambda(x^{2},y^{2},z^{2})$ = $x^{4}+y^{4}+z^{4}-2(x^{2}y^{2}+y^{2}z^{2}+z^{2}x^{2})$ is the K$\ddot{a}$llen function. $m_{c}$ and $m_{e}$ are  masses of the product mesons. $m_{q}$ and q are the mass and three momentum of the exchange meson respectively, here, pion. The decay width for the molecular state (meson and anti mesons as constituent) mediated by pion have been calculated. The expression in (13) is for the decaying via t-channel. The parameter $\Lambda_{1}$ is an adjustable constant which models the off-shell effects at the vertices due to the internal structure of the meson. Here, we assume the $\Lambda_{1}$ related to the mass of the dimesonic state with relation  $\Lambda_{1}=\kappa m_{m}$ where $\kappa$ is the constant. Hence, $\Lambda_{1}$ is depends on the dimesonic mass and fitting constant $\kappa$   \\

(2) The partial decay width $\Gamma_{j}$ (sensitive to long distance) is given by \cite{Feng-Kun Guo,Miguel}\\
$d+b\rightarrow M \rightarrow e+a+d$\\
 
Here, M is the mesonic molecule (M=d+b, where d and b are constituents) decaying into product 
mesons or photon (a,d,e).

(i) For hadronic decay: one of the constituent meson decay into two mesons
\begin{eqnarray}
\Gamma_{j}&&= \frac{1}{256 \pi^{3} m_{m}^{3}} \times \nonumber \\ 
&&\int_{({m_{a}+m_{\pi}})^{2}}^{({m_{m}-m_{a}})^{2}}dm_{ea}^{2} \int_{{(m_{ad}^{2})_{min}}}^{{(m_{ad}^{2})_{max}}} dm_{ad}^{2}|{\cal{T}}_{j}|^{2}
\end{eqnarray}

(ii) For radiative decay: one of the constituent meson decay into photon-mesons pair

\begin{eqnarray}
\Gamma_{k}&&= \frac{1}{256 \pi^{3} m_{m}^{3}} \times \nonumber \\ &&\int_{({m_{a}+m_{\pi}})^{2}}^{({m_{m}-m_{a}})^{2}}dm_{ad}^{2} \int_{{(m_{de}^{2})_{min}}}^{{(m_{de}^{2})_{max}}}dm_{de}^{2} |{\cal{T}}_{k}|^{2}
\end{eqnarray}

Here, $m_{d}$, $m_{b}$ are the constituents masses and $m_{m}$ is the mass of the dimesonic state. $g$=0.69  is the pion-meson coupling constant while $g_{1}$ is the coupling constant of mesonic molecule to the charge or neutral channels. Whereas, $|{\cal{T}}_{j}| $and $|{\cal{T}}_{k}|$ are the Feynman  amplitude for hadronic and radiative decay respectively. See the Appendix-C as well as Ref.\cite{Feng-Kun Guo,Miguel} for more detailed of the Eq.(14) and Eq(15).  
\begin{figure}[b]
\includegraphics[scale=1.0]{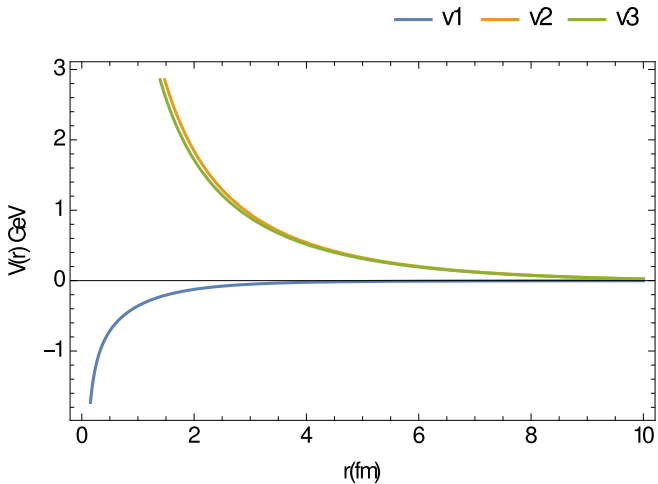}
\caption{the shape of the potential for the deuteron. The legends $V_{1}$, $V_{2}$ and $V_{3}$ shows the OPE, Hellmann and net potential, with $B_{0}$=6.5, $C_{0}$=0.235, and $\Lambda$=1.5 GeV}
\end{figure}

\section{The Deuteron from the model}
The deuteron is uncontroversially excepted as a bound state of the proton and neutron.  we tested the model on deuteron and apply to other calculations. We have fitted our potential parameters to get approximate binding energy of the deuteron. For that, The constant of the Hellmann potential are fitted as, $B_{0}$=6.5 and $C_{0}$=0.235. Whereas, the range parameter $\Lambda$, introduced in the one pion exchange potential is taken as $\Lambda$=1.5 GeV. The Fig.(1) depicted the nature of the potential for deuteron. The binding energy from the calculation to be found\\

 $BE_{(cal)}=2.856$  MeV \hspace*{0.8cm} $ \left\langle r^{2}\right\rangle_{(cal.)} =2.78$ fm \\
 
 $BE_{(Exp.)}=2.224$ MeV \hspace*{0.8cm} $ \left\langle r^{2}\right\rangle_{(cal.)} =2.14$ fm \\
 
The calculated binding energy and root mean square radius are comparable with expected experimental values. Thus, we have fixed the values of the $B_{0}$, $C_{0}$ and $\Lambda$ for rest of the calculations of the dimesonic states.
For all dimesonic sates, the constants of the Hellmann potential can be calculated by using the Eq.(6) and $\Lambda$=1.5 GeV for OPEP whereas the masses of the mesons used in the whole calculation are taken form the Particle Data Group \cite{Olive}, also tabulated in Table-I.\\

In our approach for dimesonic calculations, we have included expansion of kinetic energy and correction to the potential. We have expand the kinetic energy term up to ${\cal{O}}(P^{6})$, in these binomial expansion, contribution of the higher order term of the momentum is negligible whereas the higher order also has poor convergence. So the usable expansion to include the relativistic effect is being up to ${\cal{O}}(P^{6})$. As the $v<<c$ the higher order effect contributes less than $1 \%$. The relativistic correction added in the potential is behave as attractive coulombic term. In our previous work \cite{Rathaud}, the contribution due to the correction have been found $\sim 25-40 \%$. Whereas, in the present study this contribution in the binding energy are found $\sim 7-18 \%$ or $1-3$ MeV which reflect  in our results of $P\overline{V}$ and $V\overline{V}$-states. \\

\begin{table*}[]

\label{PV-satates}
\caption{Mass Spectra of P$\overline{V}$-states, $J^{P}$-$1^{+}$ with C-parity = $\pm$, A-without correction, B-with correction}

\begin{tabular}{ccccccccc}
\hline
System &$I^{G}(J^{PC})$ & $\mu$ & R(0)&$\sqrt{r}$ &\multicolumn{2}{c}{B.E (in MeV)} & \multicolumn{2}{c}{Mass (in GeV)} \tabularnewline
\cline{6-7} \cline{8-9}
& & $(GeV)$ &$(GeV^{\frac{3}{2}})$ & $(fm)$ & A & B & A & B \tabularnewline

\hline
$D^{0}-\overline{D^{*0}}$& $0^{-}$ $(1^{++})$ & 0.1147 & 0.0274 & 5.95 &-10.69 & -11.36 & 3.861 & 3.860  \tabularnewline

$D^{0}-\overline{D^{*0}}$& $0^{-}$ $(1^{+-})$ & 0.1141 & 0.0272 & 5.98 & -10.66 & -11.32 & 3.861 & 3.860   \tabularnewline

$D^{0}-\overline{D^{*0}}$& $1^{+}$ $(1^{++})$ & 0.1117 & 0.0264 & 6.11 & -10.45 & -11.09 & 3.861 & 3.860 \tabularnewline

$D^{0}-\overline{D^{*0}}$& $1^{+}$ $(1^{+-})$ & 0.1119 & 0.0264 & 6.10 & -10.46 & -11.10 & 3.861 & 3.860 \tabularnewline

$D^{\pm}-\overline{D^{*0}}$ & $0^{-}$ $(1^{++})$ & 0.1146 & 0.0274 & 5.96 & -10.68 & -11.35 & 3.865 & 3.865  \tabularnewline

$D^{\pm}-\overline{D^{*0}}$& $0^{-}$ $(1^{+-})$ & 0.1141 & 0.0272 & 5.98 & -10.65 & -11.31 & 3.865 & 3.865 \tabularnewline

$D^{\pm}-\overline{D^{*0}}$ & $1^{+}$ $(1^{++})$ & 0.1117 & 0.0264 & 6.11 & -10.44 & -11.08 & 3.866 & 3.865 \tabularnewline

$D^{\pm}-\overline{D^{*0}}$& $1^{+}$ $(1^{+-})$ & 0.1118 & 0.0264 & 6.10 & -10.46 & -11.10 & 3.866 & 3.865 \tabularnewline

$D^{0}-\overline{D^{*\pm}}$ & $0^{-}$ $(1^{++})$ & 0.1146 & 0.0274 & 5.96 & -10.69 & -11.36 & 3.864 & 3.863  \tabularnewline

$D^{0}-\overline{D^{*\pm}}$& $0^{-}$ $(1^{+-})$ & 0.1141 & 0.0272 & 5.98 & -10.65 & -11.32 & 3.864 & 3.863 \tabularnewline

$D^{0}-\overline{D^{*\pm}}$ & $1^{+}$ $(1^{++})$ & 0.1117 & 0.0264 & 6.11 & -10.45 & -11.09 & 3.864 & 3.864  \tabularnewline

$D^{0}-\overline{D^{*\pm}}$& $1^{+}$ $(1^{+-})$ & 0.1119 & 0.0264 & 6.10 & -10.46 & -11.10 & 3.864 & 3.864 \tabularnewline

$D^{\pm}-\overline{D^{*\pm}}$ & $0^{-}$ $(1^{++})$ & 0.1146 & 0.0274 & 5.96 & -10.68 & -11.35 & 3.869 & 3.868 \tabularnewline

$D^{\pm}-\overline{D^{*\pm}}$& $0^{-}$ $(1^{+-})$ & 0.1141 & 0.0272 & 5.99 & -10.65 & -11.31 & 3.869 & 3.868 \tabularnewline

$D^{\pm}-\overline{D^{*\pm}}$ & $1^{+}$ $(1^{++})$ & 0.1117 & 0.0263 & 6.12 & -10.44 & -11.08 & 3.869 & 3.868  \tabularnewline

$D^{\pm}-\overline{D^{*\pm}}$& $1^{+}$ $(1^{+-})$ & 0.1118 & 0.0264 & 6.11 & -10.45 & -11.09 & 3.869 & 3.868 \tabularnewline

$D^{0}-\overline{B^{*}}$& $0^{-}$ $(1^{++})$ & 0.3462 & 0.1440 & 1.97 & -25.97 & -29.11 & 7.164 & 7.160  \tabularnewline

$D^{0}-\overline{B^{*}}$& $0^{-}$ $(1^{+-})$ & 0.3414 & 0.1410 & 2.00 & -25.73 & -28.79 & 7.164 & 7.161 \tabularnewline

$D^{0}-\overline{B^{*}}$& $1^{+}$ $(1^{++})$ & 0.3323 & 0.1354 & 2.05 & -25.06 & -27.97 & 7.164 & 7.162 \tabularnewline

$D^{0}-\overline{B^{*}}$& $1^{+}$ $(1^{+-})$ & 0.3338 & 0.1363 & 2.04 & -25.14 & -28.07 & 7.164 & 7.161 \tabularnewline

$D^{\pm}-\overline{B^{*}}$ & $0^{-}$ $(1^{++})$ & 0.3460 & 0.1439 & 1.97 & -25.95 & -29.08 & 7.168 & 7.165  \tabularnewline

$D^{\pm}-\overline{B^{*}}$ & $0^{-}$ $(1^{+-})$ & 0.3413 & 0.1410 & 2.00 & -25.71 & -28.76 & 7.169 & 7.166  \tabularnewline

$D^{\pm}-\overline{B^{*}}$ & $1^{+}$ $(1^{++})$ & 0.3322 & 0.1354 & 2.05 & -25.05 & -27.94 & 7.169 & 7.166 \tabularnewline

$D^{\pm}-\overline{B^{*}}$ & $1^{+}$ $(1^{+-})$ & 0.3336 & 0.1362 & 2.04 & -25.12 & -28.04 & 7.169 & 7.166 \tabularnewline

$B^{0}-\overline{D^{*0}}$ & $0^{-}$ $(1^{++})$ & 0.3420 & 0.1414 & 1.99 & -25.46 & -28.29 & 7.261 & 7.258 \tabularnewline

$B^{0}-\overline{D^{*0}}$ & $0^{-}$ $(1^{+-})$ & 0.3375 & 0.1386 & 2.02 & -25.23 & -27.99 & 7.261 & 7.258 \tabularnewline

$B^{0}-\overline{D^{*0}}$ & $1^{+}$ $(1^{++})$ & 0.3287 & 0.1332 & 2.07 & -24.58 & -27.21 & 7.261 & 7.259  \tabularnewline

$B^{0}-\overline{D^{*0}}$ & $1^{+}$ $(1^{+-})$ & 0.3300 & 0.1341 & 2.07 & -24.65 & -27.30 & 7.261 & 7.259  \tabularnewline

$B^{0}-\overline{D^{*\pm}}$ & $0^{-}$ $(1^{++})$ & 0.3419 & 0.1413 & 1.99 & -25.45 & -28.27 & 7.264 & 7.261 \tabularnewline

$B^{0}-\overline{D^{*\pm}}$ & $0^{-}$ $(1^{+-})$ & 0.3374 & 0.1385 & 2.02 & -25.22 & -27.97 & 7.264 & 7.261 \tabularnewline

$B^{0}-\overline{D^{*\pm}}$ & $1^{+}$ $(1^{++})$ & 0.3286 & 0.1332 & 2.08 & -24.57 & -27.19 & 7.265 & 7.262  \tabularnewline

$B^{0}-\overline{D^{*\pm}}$ & $1^{+}$ $(1^{+-})$ & 0.3300 & 0.1340 & 2.07 & -24.64 & -27.28 & 7.265 & 7.262 \tabularnewline

$B^{\pm}-\overline{D^{*0}}$ & $0^{-}$ $(1^{++})$ & 0.3420 & 0.1414 & 1.99 & -25.46 & -28.29 & 7.260 & 7.257 \tabularnewline

$B^{\pm}-\overline{D^{*0}}$ & $0^{-}$ $(1^{+-})$ & 0.3375 & 0.1386 & 2.02 & -25.23 & -27.99 & 7.261 & 7.258 \tabularnewline

$B^{\pm}-\overline{D^{*0}}$ & $1^{+}$ $(1^{++})$ & 0.3287 & 0.1332 & 2.07 & -24.58 & -27.21 & 7.261 & 7.259  \tabularnewline

$B^{\pm}-\overline{D^{*0}}$ & $1^{+}$ $(1^{+-})$ & 0.3301 & 0.1341 & 2.07 & -24.65 & -27.30 & 7.261 & 7.258  \tabularnewline

$B^{\pm}-\overline{D^{*\pm}}$ & $0^{-}$ $(1^{++})$ & 0.3419 & 0.1413 & 1.99 & -25.45 & -28.27 & 7.264 & 7.261 \tabularnewline

$B^{\pm}-\overline{D^{*\pm}}$ & $0^{-}$ $(1^{+-})$ & 0.3374 & 0.1385 & 2.02 & -25.22 & -27.97 & 7.264 & 7.261 \tabularnewline

$B^{\pm}-\overline{D^{*\pm}}$ & $1^{+}$ $(1^{++})$ & 0.3286 & 0.1332 & 2.08 & -24.57 & -27.19 & 7.264 & 7.262  \tabularnewline

$B^{\pm}-\overline{D^{*\pm}}$ & $1^{+}$ $(1^{+-})$ & 0.3300 & 0.1340 & 2.07 & -24.64 & -27.28 & 7.264 & 7.262 \tabularnewline

$B^{0}-\overline{B^{*}}$ & $0^{-}$ $(1^{++})$ & 0.6140 & 0.3402 & 1.11 & -37.50 & -40.88 & 10.567 & 10.563  \tabularnewline

$B^{0}-\overline{B^{*}}$ & $0^{-}$ $(1^{+-})$ & 0.6024 & 0.3306 & 1.13 &-36.96 & -40.22 & 10.567 & 10.564  \tabularnewline

$B^{0}-\overline{B^{*}}$ & $1^{+}$ $(1^{++})$ & 0.5880 & 0.3188 & 1.16 &-35.98 & -39.09 & 10.568 & 10.565 \tabularnewline

$B^{0}-\overline{B^{*}}$ & $1^{+}$ $(1^{+-})$ & 0.5916 & 0.3218 & 1.15 & -36.14 & -39.29 & 10.568 & 10.565 \tabularnewline

$B^{\pm}-\overline{B^{*}}$ & $0^{-}$ $(1^{++})$ & 0.6140 & 0.3402 & 1.11 &-37.50 & -40.88 & 10.566 & 10.563	 \tabularnewline

$B^{\pm}-\overline{B^{*}}$ & $0^{-}$ $(1^{+-})$ & 0.6024 & 0.3306 & 1.13 & -36.96 & -40.22 & 10.567 & 10.564 \tabularnewline

$B^{\pm}-\overline{B^{*}}$ & $1^{+}$ $(1^{++})$ & 0.5880 & 0.3188 & 1.16 & -35.98 & -39.09 & 10.568 & 10.565 \tabularnewline

$B^{\pm}-\overline{B^{*}}$ & $1^{+}$ $(1^{+-})$ & 0.5916 & 0.3218 & 1.15 & -36.14 & -39.29 & 10.568 & 10.565 \tabularnewline

\hline
\end{tabular}
\end{table*}

\section{The $\bf{P\overline{V}-States}$}
The $P\overline{V}$ bound state, with a pseudoscalar meson (P=$0^{-+}$) and vector meson (V=$1^{--}$), mesons made of light quarks q (u,d,s) and heavy quarks Q(c,b).
With this definition the $P\overline{V}$ bound state of the charge conjugation parity (C) eigenstate are
\begin{eqnarray*}
C|P\overline{V}\pm \overline{P}V\rangle = \pm |P\overline{V}\pm \overline{P}V\rangle
\end{eqnarray*}
as this convention have been used by Ref\cite{Tornqvist-Z,Thomas}. Hence discussed in the section-II for the calculation of spin-isospin factor in central potential of OPEP and following the Ref. \cite{Thomas}, we have calculated the mass spectra and decay properties of the $P\overline{V}$ states. The results are tabulated in Table-II and Table-IV. We have attempted both possible choice of charge conjugation parity (C=$\pm$) for the $P\overline{V}$ states calculations, which impact on the sign of the spin-isospin factors according to Eq(8). The Table-II depicted the mass spectra of possible $P\overline{V}$ states with heavy-light flavour mesons. From the analysis of the calculated results, we have found the very small but noticeable effect of charge conjugation parity and isospin channels on the binding energy results. \\

In the same isospin space, with different charge conjugation parity, the change in the binding energy is $\sim 0.03$ MeV (with total isospin I=0) and  $\sim 0.01$ MeV (with total isospin I=1), for both charge or neutral channels of dimesonic combinations. For total isospin I=0, the binding energy is more for C=$+$ then C=$-$, it implies, the binding energy decreases as we change the sign of C from +1 to -1. But, With total isospin I=1, the binding energy increases with change the sign of C from +1 to -1, opposite to the I=0 channel. It clearly shows the symmetry breaking in same isospin space. Moreover, with different isospin space and same charge conjugation parity, the change in the binding energy is  $\sim 0.22$ MeV for both charge and neutral channels. The same effect is shown with different isospin and C parity. The effect is also multiplicatively increases as mass of the system increases. In Ref.\cite{Tornqvist-L}, T$\ddot{o}$rnqvist had discussed about the isospin symmetry breaking and same was mentioned in \cite{Thomas}.\\

\subsection{The $D\overline{D^{*}}$, $B\overline{B^{*}}$, $D\overline{B^{*}}$ as dimesonic molecules}

The bound state of $D\overline{D^{*}}$ and $B\overline{B^{*}}$ have studied since long time. In the various literature\cite{Tornqvist-Z,Thomas,Gui-Jun Ding,F. E. Close,Miguel,Feng-Kun Guo,Swanson}, there have been discussed about the possibility of the state X(3872) as a $D\overline{D^{*}}$ molecule as well as predicted the possible partner of $D\overline{D^{*}}$ in the bottomonium sector as a $B\overline{B^{*}}$ bound state. We  have calculated the binding energy and masses of the $D\overline{D^{*}}$ and $B\overline{B^{*}}$ dimesonic states (See Table-II). We have attempted all possible combination of charge and neutral charm and bottom meson for dimesonic calculations. For the state X(3872) as $D\overline{D^{*}}$ molecule, the binding energy and mass to be found
\begin{eqnarray}
BE_{D\overline{D^{*}}} \cong  10.56 MeV ; M_{D\overline{D^{*}}} \cong  3.865 GeV \nonumber
\end{eqnarray}
the binding energy and mass of $B\overline{B^{*}}$ dimesonic state to be found 
\begin{eqnarray}
BE_{B\overline{B^{*}}} \cong  36.80 MeV ; M_{B\overline{B^{*}}} \cong  10.564 GeV \nonumber
\end{eqnarray}
The binding energy are overestimated due to the dominance of the Hellmann potential. The dominating quantum number for the state X(3872) is $1^{+}$ with positive charge parity C=+ and total isospin I=0 \cite{A. Abulencia}. In our study, the binding energy of the dimesonic states $D^{0}\overline{D^{*0}}$, $D^{\pm}\overline{D^{*0}}$,$D^{0}\overline{D^{*\pm}}$ and $D^{\pm}\overline{D^{*\pm}}$  with total isospin I=0 are almost equal, whereas, with I=1 these states are less bound compare to I=0. Thus, we have also suggested $J^{P}=1^{+}$ with total isospin I=0 and positive charge parity C=+ for X(3872) and further calculating the decay properties accordingly.\\

In some literature \cite{Gui-Jun Ding,Mahiko Suzuki,Yan-Rui Liu,N.N. Achasov}, the X(3872) is predicted not as a pure $D^{0}\overline{D^{*0}}$ but it acquired the coupling with its charge channels or charmonium hybrid. Here, we have not included  any coupling scheme in our calculations such that we can not reach on any conclusion from mass spectra that X(3872) is a pure $D^{0}\overline{D^{*0}}$ molecule or has some mixing of the channels. For such testing one have to study the decay properties of the state. So we have calculated some partial decay width as per discussed in section-2 to probe some conclusion on the internal structure of the state X(3872). In a straight way, to check the molecular picture of the X(3872), we just probe the decay process which sensitive to its long and short distance structure and identified its dominant decay mode.\\
By using the Eq(14), we have calculated the decay mode sensitive to the long distance structure. The decay process could be expressed as
\begin{eqnarray}
D\overline{D^{*}} \longrightarrow D\overline{D}\pi^{0} \nonumber
\end{eqnarray}
Thus the partial decay width we have
\begin{eqnarray}
\Gamma_{D\overline{D^{*}}\rightarrow D\overline{D}\pi^{0}}=0.0455 MeV \nonumber
\end{eqnarray}
While by using the Eq(12), we have calculated the decay mode sensitive to the short distance structure. The decay process could be expressed as
\begin{eqnarray}
D\overline{D^{*}}&& \longrightarrow J/\psi \rho \longrightarrow J/\psi\pi^{+}\pi^{-} \nonumber \\ \nonumber
D\overline{D^{*}}&& \longrightarrow J/\psi \omega \longrightarrow J/\psi\pi^{+}\pi^{-}\pi^{0} \\ \nonumber
D\overline{D^{*}}&& \longrightarrow J/\psi \gamma  \\ \nonumber
D\overline{D^{*}}&& \longrightarrow \psi(2S) \gamma  \\ \nonumber
\end{eqnarray}    
Where we assume as in Ref.\cite{P. del Amo,Kunihiko}, the $D\overline{D^{*}}$ decay to the final state decay product $ J/\psi 2\pi$ and  $J/\psi 3\pi$  via decay of $\rho$ and $\omega$ respectively as 
\begin{eqnarray}
\rho && \longrightarrow \pi^{+}\pi^{-} \nonumber \\ 
\omega && \longrightarrow \pi^{+}\pi^{-}\pi^{0} \nonumber
\end{eqnarray} 
Thus the partial decay width (with ${\cal{K}}=0.991$),  we have
\begin{eqnarray}
\Gamma_{D\overline{D^{*}}\rightarrow J/\psi 2\pi}&=&0.5226 MeV \nonumber \\
\Gamma_{D\overline{D^{*}}\rightarrow J/\psi 3\pi}&=&0.4769 MeV  \nonumber \\
\Gamma_{D\overline{D^{*}}\rightarrow J/\psi \gamma}&=&0.5633 MeV  \nonumber \\
\Gamma_{D\overline{D^{*}}\rightarrow \psi(2S) \gamma}&=&0.1392 MeV  \nonumber
\end{eqnarray}

Thus the total width could be calculated as sum of all partial width.  We have 
 
\begin{eqnarray}
\Gamma_{D\overline{D^{*}}}&&= 1.74 MeV \nonumber \\
\Gamma_{D\overline{D^{*}}_(Exp.)} && < 1.2 MeV  \nonumber
\end{eqnarray}

The total decay width is overestimated to the experimental value. In Ref. \cite{Aaij} LHCb  Collaboration presented the evidence for the  decay mode $\Gamma_{(X(3872)}\rightarrow \psi(2S)\gamma$ and  the ratio of the branching fraction for the mode measured to be $2.4\pm0.64\pm0.29$ and supported the previous  experimental results of the Belle \cite{Bhardwaj} and BABAR \cite{Aubert}. 
This results excluded the possible interpretation of the pure molecular picture of the X(3872) and agrees with the interpretation as a pure charmonium  or dominant molecular-charmonium mixture.  In our calculation the the ratio of the branching fraction to be found 0.24 which is ten times less than the reported experimental results. Moreover, the values of the radiative decay rate to be found large deviation with the other theoretical studies \cite{Mehen,Guo,Cincioglu,A. M. Badalian,V. D. Orlovsky,Barnes,T. Barnes,Y. Dong}. In Ref. \cite{Guo} argued that the radiative decay and their ratio are very weakly sensitive to the long range structure of the X(3872) and thus they can not be used to rule out the molecular interpretation of the state.  One need to probe the decay modes sensitive to the long distance structure of the state. Hence, we have attempted both decay modes and calculated results are partially in agreements with experimental as well as theoretical studies.       
However, in the same manner, we have calculated decay properties for the $B\overline{B^{*}}$ bound state as a possible partner of  $D\overline{D^{*}}$ molecule. For the decay mode sensitive to the short distance structure, the decay processes expressed as 
\begin{eqnarray*}
B\overline{B^{*}} &&\longrightarrow \Upsilon(1S) \pi^{+} \pi^{-} \nonumber \\
B\overline{B^{*}} &&\longrightarrow \Upsilon(2S) \pi^{+} \pi^{-}
\end{eqnarray*}
Thus the calculated partial decay widths (with ${\cal{K}}=0.998$), we have 
\begin{eqnarray*}
\Gamma_{B\overline{B^{*}}\rightarrow \Upsilon(1S) \pi^{+} \pi^{-}}&=&19.843 MeV \nonumber \\
\Gamma_{B\overline{B^{*}}\rightarrow \Upsilon(2S) \pi^{+} \pi^{-}}&=&8.923 MeV
\end{eqnarray*} 

The vales are comparable to the Ref. \cite{Bonder} and \cite{J. M. Dias,Bonder-2} predicted the $Z_{b}(10604)$ as $B\overline{B^{*}}$  molecule. In our model, the decay modes sensitive to the long distance structure of $B\overline{B^{*}}$ molecule to be found forbidden. The calculated decay properties are tabulated in Table-IV.\\
 
\begin{table*}[]
\caption{Mass Spectra of V$\overline{V}$-states, with $I$ $(J^{P})$,  A-without correction, B-with }

\begin{tabular}{ccccccccc}
\hline
System &$I^{G}(J^{PC})$ & $\mu$ & R(0)&$\sqrt{r^{2}}$ &\multicolumn{2}{c}{B.E (in MeV)} & \multicolumn{2}{c}{Mass (in GeV)} \tabularnewline
\cline{6-7} \cline{8-9}
& & $(GeV)$ &$(GeV^{\frac{3}{2}})$ & $(fm)$ & A & B & A & B  \tabularnewline
\hline
$D^{*0}-\overline{D^{*0}}$& $0^{}$ $(0^{+})$ & 0.1120 & 0.0265  & 6.10 & -10.38 & -10.98 & 4.003 & 4.002  \tabularnewline
$D^{*0}-\overline{D^{*0}}$& $0^{}$ $(1^{+})$ & 0.1096 & 0.0256 & 6.23 & -10.17 & -10.75 & 4.003 & 4.003  \tabularnewline
$D^{*0}-\overline{D^{*0}}$& $0^{}$ $(2^{+})$ & 0.1107  & 0.0260 & 6.17 & -10.26 & -10.85 & 4.003 & 4.003  \tabularnewline
$D^{*0}-\overline{D^{*0}}$& $1^{}$ $(0^{+})$ & 0.1114 & 0.0263 & 6.13 & -10.32 & -10.91 & 4.003 & 4.002  \tabularnewline
$D^{*0}-\overline{D^{*0}}$& $1^{}$ $(1^{+})$ & 0.1122 & 0.0265 & 6.08 & -10.39 & -10.99 & 4.003 & 4.002  \tabularnewline
$D^{*0}-\overline{D^{*0}}$& $1^{}$ $(2^{+})$ & 0.1118 & 0.0264 & 6.11 & -10.35 & -10.96 & 4.003 & 4.003  \tabularnewline

$D^{*0}-\overline{D^{*\pm}}$& $0^{}$ $(0^{+})$ & 0.1119 & 0.0265 & 6.10 & -10.37 & -10.97 & 4.006 & 4.006  \tabularnewline
$D^{*0}-\overline{D^{*\pm}}$& $0^{}$ $(1^{+})$ & 0.1096 & 0.0256 & 6.23 & -10.17 & -10.75 & 4.007 & 4.006  \tabularnewline
$D^{*0}-\overline{D^{*\pm}}$& $0^{}$ $(2^{+})$ & 0.1107 & 0.0260 & 6.17 & -10.25 & -10.84 & 4.007 & 4.006  \tabularnewline
$D^{*0}-\overline{D^{*\pm}}$& $1^{}$ $(0^{+})$ & 0.1114 & 0.0263 & 6.13 & -10.31 & -10.91 & 4.006 & 4.006  \tabularnewline
$D^{*0}-\overline{D^{*\pm}}$& $1^{}$ $(1^{+})$ & 0.1122 & 0.0265 & 6.08 & -10.38 & -10.99 & 4.006 & 4.006  \tabularnewline
$D^{*0}-\overline{D^{*\pm}}$& $1^{}$ $(2^{+})$ & 0.1118 & 0.0264 & 6.11 & -10.35 & -10.95 & 4.006 & 4.006  \tabularnewline

$D^{*\pm}-\overline{D^{*\pm}}$& $0^{}$ $(0^{+})$ & 0.1119 & 0.0264 & 6.10 & -10.37 & -10.97 & 4.010 & 4.009  \tabularnewline
$D^{*\pm}-\overline{D^{*\pm}}$& $0^{}$ $(1^{+})$ & 0.1096 & 0.0256 & 6.23 & -10.16 & -10.74 & 4.010 & 4.009  \tabularnewline
$D^{*\pm}-\overline{D^{*\pm}}$& $0^{}$ $(2^{+})$ & 0.1107 & 0.0260 & 6.17 & -10.25 & -10.84 & 4.010 & 4.009  \tabularnewline
$D^{*\pm}-\overline{D^{*\pm}}$& $1^{}$ $(0^{+})$ & 0.1114 & 0.0262 & 6.13 & -10.31 & -10.90 & 4.010 & 4.009  \tabularnewline
$D^{*\pm}-\overline{D^{*\pm}}$& $1^{}$ $(1^{+})$ & 0.1122 & 0.0265 & 6.09 & -10.38 & -10.98 & 4.010 & 4.009  \tabularnewline
$D^{*\pm}-\overline{D^{*\pm}}$& $1^{}$ $(2^{+})$ & 0.1118 & 0.0264 & 6.11 & -10.36 & -10.96 & 4.010 & 4.009  \tabularnewline

$D^{*0}-\overline{B^{*}}$ & $0^{}$ $(0^{+})$ & 0.3313 & 0.1348 & 2.06 & -24.82 & -27.48 & 7.306 & 7.303  \tabularnewline
$D^{*0}-\overline{B^{*}}$ & $0^{}$ $(1^{+})$ & 0.3224 & 0.1294 & 2.12 & -24.16 & -26.68 & 7.307 & 7.305  \tabularnewline
$D^{*0}-\overline{B^{*}}$ & $0^{}$ $(2^{+})$ & 0.3294 & 0.1336 & 2.07 & -24.57 & -27.19 & 7.307 & 7.305  \tabularnewline
$D^{*0}-\overline{B^{*}}$ & $1^{}$ $(0^{+})$ & 0.3318 & 0.1351 & 2.06 & -24.76 & -27.42 & 7.306 & 7.303  \tabularnewline
$D^{*0}-\overline{B^{*}}$ & $1^{}$ $(1^{+})$ & 0.3350 & 0.1371 & 2.04 & -24.99 & 27.70 & 7.306 & 7.304  \tabularnewline
$D^{*0}-\overline{B^{*}}$ & $1^{}$ $(2^{+})$ & 0.3325 & 0.1355 & 2.05 & -24.84 & -27.51 & 7.307 & 7.305  \tabularnewline

$D^{*\pm}-\overline{B^{*}}$ & $0^{}$ $(0^{+})$ & 0.3312 & 0.1347 & 2.06 & -24.81 & -27.46 & 7.310 & 7.307  \tabularnewline
$D^{*\pm}-\overline{B^{*}}$ & $0^{}$ $(1^{+})$ & 0.3223 & 0.1293 & 2.12 & -24.15 & -26.66 & 7.311 & 7.308  \tabularnewline
$D^{*\pm}-\overline{B^{*}}$ & $0^{}$ $(2^{+})$ & 0.3293 & 0.1336 & 2.07 & -24.56 & -27.18 & 7.311 & 7.308  \tabularnewline
$D^{*\pm}-\overline{B^{*}}$ & $1^{}$ $(0^{+})$ & 0.3317 & 0.1351 & 2.06 & -24.74 & -27.40 & 7.310 & 7.307  \tabularnewline
$D^{*\pm}-\overline{B^{*}}$ & $1^{}$ $(1^{+})$ & 0.3349 & 0.1370 & 2.04 & -24.97 & -27.68 & 7.310 & 7.307  \tabularnewline
$D^{*\pm}-\overline{B^{*}}$ & $1^{}$ $(2^{+})$ & 0.3324 & 0.1355 & 2.05 & -24.83 & -27.50 & 7.310 & 7.308  \tabularnewline
$B^{*}-\overline{B^{*}}$ & $0^{}$ $(0^{+})$ & 0.5895 & 0.320 & 1.15 & -36.21 & -39.32 & 10.613 & 10.610  \tabularnewline
$B^{*}-\overline{B^{*}}$ & $0^{}$ $(1^{+})$ & 0.5753 & 0.308 & 1.18 & -35.21 & -38.18 & 10.614 & 10.611  \tabularnewline
$B^{*}-\overline{B^{*}}$ & $0^{}$ $(2^{+})$ & 0.5912 & 0.321 & 1.15 & -36.03 & -39.15 & 10.615 & 10.612  \tabularnewline
$B^{*}-\overline{B^{*}}$ & $1^{}$ $(0^{+})$ & 0.5951 & 0.324 & 1.14 & -36.31 & -39.47 & 10.612 & 10.609  \tabularnewline
$B^{*}-\overline{B^{*}}$ & $1^{}$ $(1^{+})$ & 0.6001 & 0.328 & 1.13 & -36.66 & 39.87 & 10.613 & 10.609  \tabularnewline
$B^{*}-\overline{B^{*}}$ & $1^{}$ $(2^{+})$ & 0.5945 & 0.324 & 1.14 & -36.37 & 39.53 & 10.614 & 10.611  \tabularnewline

\hline
\end{tabular}
\end{table*}

\section{The $\bf{V\overline{V}-States}$}
The vector meson V=$1^{--}$, defining the meson and antimeson state as 
\begin{eqnarray*}
V&=& \frac{1}{\sqrt{2}}[q(1)\overline{Q}(2)-\overline{Q}(1)q(2)] \nonumber \\
\overline{V}&=&  \frac{1}{\sqrt{2}}[\overline{q}(1)Q(2)-Q(1)\overline{q}(2)]
\end{eqnarray*} 
as the convention used in the Ref\cite{Thomas}. The spin-isospin factor for $V\overline{V}$ states can be calculated as discussed in section-II. The  $V\overline{V}$ states are spin triplet. So we have performed calculations for $V\overline{V}$ with possible spin-isospin space. The calculated results for mass spectra are tabulated in Table-III. The strength of the different spin-isospin channels according to our calculation are discussed in appendix-B(also see Fig-3). In the literature, it have been predicted the possible $D^{*}\overline{D^{*}}$ and $B^{*}\overline{B^{*}}$ bound states with the study of $D\overline{D^{*}}$ and $B^{*}\overline{B^{*}}$ molecules. In Ref.\cite{Miguel}, authors investigate and suggested spin partner to $D\overline{D^{*}}$ with quantum number $2^{++}$, according to heavy quark spin symmetry. It would be a $D^{*}\overline{D^{*}}$ bound state, $X_{c2}$. Similarly, one could expect bottom partner with quantum number $2^{++}$, would be $B^{*}\overline{B^{*}}$ bound state, $X_{b2}$.  The calculated binding energy and mass of the $X_{c2}$ state 
\begin{eqnarray*}
BE_{D^{*}\overline{D^{*}}}&=&X_{c2}\cong  10.27 MeV \nonumber \\
M_{D^{*}\overline{D^{*}}}&=&X_{c2} \cong  4.006 GeV
\end{eqnarray*}
Whereas the  binding energy and mass of the $X_{b2}$ state being
\begin{eqnarray*}
BE_{B^{*}\overline{B^{*}}}&=&X_{b2}\cong  35.93 MeV \nonumber \\
M_{B^{*}\overline{B^{*}}}&=&X_{b2} \cong  10.613 GeV
\end{eqnarray*}

\begin{table*}
\caption{Decay Width of dimesonic states}
\begin{tabular}{cccccc}
\hline
Dimesonic & Decay mode & Partial & others & Total & State \tabularnewline
State & & Width (in MeV) & & Width & \tabularnewline
\hline
$D\overline{D^{*}}$ & $X_{c}\rightarrow J/\psi 2\pi$ &0.522 & & & \tabularnewline
          & $X_{c}\rightarrow J/\psi 3\pi$ &0.476 & &  & \tabularnewline
          & $X_{c}\rightarrow J/\psi \gamma$ &0.563 &0.05\cite{V. D. Orlovsky} & 1.74 & $X(3872)$\tabularnewline
           & $X_{c}\rightarrow \psi(2S) \gamma$ &0.139 &0.180\cite{T. Barnes} &  & \tabularnewline
          & $X_{c}\rightarrow D \overline{D} \pi$ &0.045 & & & \tabularnewline
\hline          
$D^{*}\overline{D^{*}}$ & $X_{c2}\rightarrow D \overline{D}$ &1.076 & $0.6_{-0.2}^{+0.7}$ \cite{Miguel}& & \tabularnewline
                        & $X_{c2}\rightarrow D \overline{D^{*}}$ &0.752 &$0.7_{-0.2}^{+0.5}$\cite{Miguel} & 1.84 &$X(4013)$ \tabularnewline
                   & $X_{c2}\rightarrow D \overline{D^{*}}\gamma$ &0.013 &$0.018_{-0.006}^{+0.002}$ \cite{Miguel}& & \tabularnewline
\hline
$B\overline{B^{*}}$ & $X_{b}\rightarrow \Upsilon(1S)\pi^{+}\pi^{-}$ & 19.84 & $22.3\pm7.7$ \cite{Bonder} & &\tabularnewline 
            & $X_{b}\rightarrow \Upsilon(2S)\pi^{+}\pi^{-}$ & 8.923 & $24.2\pm3.1$ \cite{Bonder}& 28.7&$X_{b}(10610)$/$Z_{b}(10610)$ \tabularnewline 
            & $X_{b}\rightarrow B \overline{B}$ & F & & & \tabularnewline   
\hline             
$B^{*}\overline{B^{*}}$ & $X_{b2}\rightarrow B \overline{B}$ &4.014 &$4.4_{-0.4}^{+0.1}$ \cite{Miguel}& & \tabularnewline
          & $X_{b2}\rightarrow B \overline{B^{*}}$ &1.415 &$2.0_{-1.0}^{+0.9}$ \cite{Miguel}&5.42 &$X_{b}(10650)$/$Z_{b}(10650)$  \tabularnewline
          & $X_{b2}\rightarrow B \overline{B^{*}}\gamma$ &F& 6 $10^{-6}$ \cite{Miguel}& & \tabularnewline
\hline
       
\end{tabular}
\end{table*}

There are large possibilities of dimesonic states with heavy light flavors, also shown in our calculated mass spectrum(see Table-III). We have predicted the mass and root mean square radius. For instant with $V\overline{V}$ dimesonic states, for the decay properties, we focus on these two states $X_{c2}$ and $X_{b2}$ as spin-2 partner of $D\overline{D^{*}}$ and $B\overline{B^{*}}$ molecules respectively and also expected as per heavy quark spin symmetry(HQSS). The study of decay properties with mass spectra is very important to investigate their sub structure. We have calculated the hadronic and radiative decay of these states. 
The hadronic decay modes of $X_{c2}$ 
\begin{eqnarray*}
{(D^{*}\overline{D^{*}})}&=&X_{c2}\longrightarrow D \overline{D} \nonumber \\
{(D^{*}\overline{D^{*}})}&=&X_{c2}\longrightarrow D \overline{D^{*}}
\end{eqnarray*}

by using the Eq(12)(with $\cal{K}$=0.993),  the calculated partial decay width, we have

\begin{eqnarray*}
\Gamma_{X_{c2}\rightarrow D\overline{D}}&=& 1.076 MeV \nonumber \\
\Gamma_{X_{c2}\rightarrow D\overline{D^{*}}}&=& 0.752 MeV
\end{eqnarray*} 
similarly the decay width for $X_{b2}$ with the same formula (with $\cal{K}$=0.998) to be found
\begin{eqnarray*}
\Gamma_{X_{b2}\rightarrow B\overline{B}}&=& 4.014 MeV \nonumber \\
\Gamma_{X_{c2}\rightarrow B\overline{B^{*}}}&=& 1.415 MeV
\end{eqnarray*} 

To understand the electromagnetic branching fraction we need to understand the interaction with photon and s-wave mesons and their contribution due to light and heavy quarks \cite{Miguel}. In addition to, radiative decay is more sensitive to long distance molecular structure. The radiative decay mode of $X_{c2}$ expressed as

\begin{eqnarray*}
X_{c2}\longrightarrow && D^{*} \overline{D}\gamma \nonumber \\
X_{c2}\longrightarrow && D^{*\pm} \overline{D^{\pm}}\gamma
\end{eqnarray*} 
thus the decay width for the state calculated as per Eq(15) (with $\Lambda_{2}$ =1 GeV), we have 

\begin{eqnarray*}
\Gamma_{X_{c2}\rightarrow \overline{D}D^{*}\gamma}&=& 13.874 KeV \nonumber \\
\Gamma_{X_{c2}\rightarrow \overline{D^{\pm}}D^{*\pm}\gamma}&=& 0.5878 KeV
\end{eqnarray*}   

The calculated partial decay widths are in good agreement with the results calculated in Ref\cite{Miguel}. The radiative decay for the state $X_{b2}$ could also calculated similar to  $X_{c2}$. 

\begin{eqnarray*}
X_{b2}\longrightarrow && B^{*} \overline{B}\gamma 
\end{eqnarray*} 

For this decay mode of $X_{b2}$, in our model the radiative decay of $X_{b2}$ to be found forbidden with suggestive value of $\Lambda_{2}$ =1 GeV. With very large value of $\Lambda_{2}$ ($\sim$8), we have get the comparable results with Ref. \cite{Miguel}. Moreover, the radiative decay width for  $X_{b2}$ calculated in  \cite{Miguel} is $\sim$ 10 eV, which is very small. With very large value of $\Lambda_{2}$, we have found  $\sim$ 14 eV. (See the Table-IV, for the results of calculated partial widths with comparisons with others).  \\

\section{Summery}
In summary, we have calculated the mass spectra of dimesonic states with heavy-light flavour mesons by using the Hellmann and one pion exchange potential. The calculated binding energy of dimesonic states are found overestimated which is also expected in variational approach. We have analyzed the change in the binding energy due to charge conjugation parity and isospin, agreed with Ref.\cite{Thomas,Tornqvist-L} suggested isospin symmetry breaking. We have also discussed the effect of tensor term in one pion exchange (see Appendix). We includes only the s-wave contribution of tensor term and observed shuffling of channels which is different from expected channels as in \cite{Tornqvist,Tornqvist-Z}. It is remarkably pointing out the effect of the tensor term and its contribution in different processes, led us to conclude that it can not be ignore. Whereas, the contribution from the relativistic correction in the potential to the binding energy is $\sim 7-18 \%$ which is almost one third to our previous work \cite{Rathaud}.\\   
    
In addition, we have calculated the decay properties of $D\overline{D^{*}}$, $B\overline{B^{*}}$, $D^{*}\overline{D^{*}}$ and $B^{*}\overline{B^{*}}$ (the formalism adopted from Ref.\cite{Rui Zhang,Feng-Kun Guo,Miguel}), using our calculated binding energy as a input where formalism mainly dependent on the masses. To validate our predictions, we have attempted the decay calculations which required to test the molecular structure of compared states with dimesonic states on the basis of decay modes which may have responsible for their long and short distance structure, as studied in the literature \cite{Miguel,Feng-Kun Guo,Swanson,F. Aceti,Eric Braaten,N.N. Achasov,Yubing Dong}. However, both the decay calculation sensitive to short and long range structure of the state are extremely sensitive to the regularization parameter (appeared in the formulas) compare to binding energy which is certainly related to the hadron size. We have found comparable results of decay properties (with our mass spectra) \cite{Bonder,Miguel,Feng-Kun Guo,Bonder-2}, support the prediction of  $D\overline{D^{*}}$, $B\overline{B^{*}}$, $D^{*}\overline{D^{*}}$ and $B^{*}\overline{B^{*}}$ bound states as mesonic molecules. In Ref. \cite{Aaij} LHCb  Collaboration presented the evidence for the  decay mode $\Gamma_{(X(3872)}\rightarrow \psi(2S)\gamma$ and this results excluded the possible interpretation of the pure molecular picture of the X(3872) and agrees with the interpretation as a pure charmonium or dominant molecular-charmonium mixture. \\

{\bf The radiative decays are important to revel the internal structure of the X(3872), such as, the decay channel X(3872) $\rightarrow J/\psi \gamma$ indicates that this state has positive C-parity. Swanson in \cite{EricSwanson} pointed out that radiative decays of the molecular X to charmonium may arise due to vector meson dominance in the $\rho J/\psi$ or $\omega J/\psi$ component of the X and led the conclusion to the dominant molecule $D\overline{D^{*}}$ plus small admixtures of $\rho J/\psi$ and $\omega J/\psi$. Thus $\gamma J/\psi$ is the only possible final state  available to  this mechanism. The decay X(3872)$\rightarrow J/\psi \pi^{+} \pi^{-}$ and X(3872) $\rightarrow J/\psi \pi^{+} \pi^{0} \pi^{-}$ through $\rho$ and $\omega$ resonances indicates large isospin breaking which could be naturally explain in the molecular model. The radiative E1 decay widths of X(3872) have been reported in  \cite{Tian-Hong Wang} where they approximated this state as a pure charmonium $\chi_{c1}(2P)$ and fitted the model parameters for $\chi_{c1}(2P)$ then further calculated the radiative decay widths, while \cite{F. De Fazio,J. Ferretti} have  also pointed the state X(3872) as a pure charmonium state $\chi_{c1}(2P)$ and \cite{F. De Fazio}have calculated the radiative decays, moreover, \cite{Yu. S. Kalashnikova} assign $^{1}D_{2}$  charmonium state to the X(3872), favoring $J^{PC}$ as $2^{-+}$ which is almost discarded in current scenario. In the review for the status of the X(3872), Suzuki \cite{Mahiko Suzuki} point out it as a charmonium state mixing with molecular $D\overline{D^{*}}+\overline{D}D^{*}$  component. In \cite{Mahiko Suzuki} author argued on binding through pion-exchange model for the pure $D\overline{D^{*}}+\overline{D}D^{*}$ molecule while this conclusion criticized in \cite{E. S. Swanson}. Braaten \cite{Eric Braaten} analyze the light meson exchange in simple model with spin-0 meson upto next-to-leading  order  in  the interaction  strength and notice that X(3872) seems to be a loosely-hadronic bound state molecule.} Indeed, in Ref. \cite{Guo} argued that the radiative decay and their ratio are very weakly sensitive to the long range structure of the X(3872) and thus they can not be used to rule out the molecular interpretation of the state.  One need to probe the decay modes sensitive to the long distance structure of the state.\\

Hence, in such a muddy scenario, we have attempted both decay modes where the calculated results are partially in agreements with experimental as well as other theoretical studies. The high sensitivity of decay calculation on the regularization parameter and lack of the other effects like coupled channel treatment in the present model restrict us to conclude the pure molecule interpretation of the state, but, still these results (except radiative decays where the ratio is underestimated almost ten times then experimental results) gives signature for the dominant molecular component carried by the state X(3872). Thus, the other possibilities may be the dominant molecular $D\overline{D^{*}}$ and admixture of the $\rho J/\psi$ and $\omega J/\psi$ as pointed by Ref. \cite{EricSwanson}. Apart from these , we have also predicted the mass spectra and decay properties of the possible spin-2 partner ($J^{PC}=2^{++}$) of X(3872) in charm sector as well as spin-2 partner of $B\overline{B^{*}}$ in the bottom sector, as Miguel et.al studied in \cite{Miguel}, also expected in heavy quark spin-flavour symmetry. Moreover, for better understanding of the molecular picture the coupled channel effect as well S-D wave mixing need to be incorporated which lack in the present study and considered as a limitation of the study. \\

Furthermore, in future we will incorporate couple channel effect and S-D wave mixing for the study of the dimesonic systems. We will also calculate the mass spectra of possible dimesonic states as mesonic molecules with $D_{s}$, $D_{s}^{*}$, $B_{s}$, $B_{s}^{*}$  mesons. As Wei Chen et.al in \cite{Wei Chen} suggested open flavour tetraquark structure with having strange quark, with mass range 6.9-7.3 GeV, with $J^{p}=0^{+} $ or $1^{+}$. In Ref.\cite{Gui-Jun Ding 2}, Gui-Jun Ding suggest $D_{s}^{*}\overline{D_{s}^{*}}$  molecule. The masses of  $P\overline{V}$ and $V\overline{V}$ dimesonic states with $D_{s}$, $D_{s}^{*}$, $B_{s}$ and $B_{s}^{*}$ mesons are fall in the same range of 7.1-7.3 GeV. We look forward to the experimental facilities and working groups to take more attention for searches of the possible dimesonic states as well as for the confirmations of our theoretical predictions.          \\

\noindent {\bf Acknowledgements}
D. P. Rathaud would like to thanks Prof. R. C. Johnson for the useful discussion. A. K. Rai acknowledge the financial support extended by D.S.T., Government of India  under SERB fast track scheme SR/FTP /PS-152/2012. 
\newpage
\appendix
{
\section{The Hellmann and One Pion Exchange Potential}

\label{sec:sect2}

Two color neutral states formed the bound states, just like deuteron (the bound state of proton and neutron). In the case of the dimesonic states, the meson and antimeson are takeover as a constituents, forming a bound state. The interaction potential between two color neutral constituents taken as a phenomenological Hellmann potential \cite{H. Hellmann} with One pion Exchange Potential(OPEP) \cite{Paris,Pandharipande}.
It is very difficult to explain such interaction at fundamental level whereas this effective potential had used previously in analogous to QED. Here, we assume that the two color neutral states (meson and antimeson) interact at very short distance in such a way that the constituent quark of the each color neutral state feels the effect of the quark of other color neutral states (just like the polar molecules). The attraction between color singlet states comes from the virtual excitation of the color octet dipole of the each hadron. Simultaneously, the color charge screening effect in this interaction occurs which is analogues to the charge screening effect as in the case of polar molecule. We assume that the overall effective potential (coulomb $+$ Yukawa) used in this study accomplished these interaction effects. Such interaction are responsible for the residual force between two color neutral states and manifestation of the energy. This manifestation of the energy may responsible  for the creation of the quark-antiquark pair. Here, the creation of the quark-antiquark pairs could represent the heavy boson exchange at short distance as studied in the case of deuteron and NN-scattering. Hence, we have replaced the complicated heavy boson exchange potential by our effective potential (Hellmann Potential), namely  
   
\begin{equation}
V_{h}{(r_{12})}=-\frac{k_{mol}}{r_{12}}+B\frac{e^{-C r_{12}}}{r_{12}}
\end{equation}

Here, the constant $k_{mol}$ is the residual strength of the strong running coupling constant between the two color neutral states and B are the strength of the Yukawa potential whereas $r_{12}$ is the relative separation between constituents. The value of the $k_{mol}$ could be determined through the model, such as
 
\begin{equation}
k_{mol}(M^{2}) = \frac{4\pi}{(11-\frac{2}{3}n_{f})ln\frac{M^{2}+ {M_{B}}^{2}}{\Lambda_{Q}^{2}}}
\end{equation}
where $m_{d}$ and $m_{b}$ are constituent masses,  M=2$m_{d}$ $m_{b}$/ ($m_{d}$+$m_{b}$), $M_{B}$=1 GeV, $\Lambda_{Q}$=0.250 GeV and $n_{f}$ is number of flavour \cite{Ebert,Badalian}. 
This effective coupling constant introduced to incorporate the asymptotic behavior at short distance as well as to reduce the free parameter of the model. This effective running coupling constant previously used in the study of mass spectra of the light mesons in the relativistic quark model \cite{Ebert}, where they adopt $\alpha_{s}\equiv \alpha_{s}(\mu^{2})$ (where $\mu^{2}=m_{1}m_{2}/m_{1}+m_{2}$) from simplest model with freezing \cite{Badalian}. \\
Now, the constant B and C appeared in the Yukawa potential of Eq.(16) play very trivial role on overall characteristic of the Hellmann potential. In the Fig(2), one can observe that for the fixed coulombic interaction, variation of the constant B and C are inversely proportional. As the value of the B increases, it increases the repulsive nature of the potential while the constant C  increases the strength of the Yukawa part-inversely to B. To study and show the nature of the Hellmann potential, the graph have been plotted for various set of values of the B and C (see the Fig(2)). For the bound state, the value of B can take both positive and negative values. With the negative values of the B the overall potential become attractive. For more detailed discussion on the Hellmann potential, we suggest to the readers for Ref.\cite{Adamowski}. In the case of the dimesonic bound states calculations, we assume that the Hellmann potential at very short distance cares for the delicate cancellation of attraction and repulsion respectively which is mainly taken care by heavy boson exchange in the One Boson Exchange model.\\

Here, the long range behavior of the interaction part is accomplished by One Pion Exchange. The model carried the net potential as the Hellmann potential plus OPE potential plus relativistic correction. For instance, to fix the model (or the potential) for deuteron which is widely believed to be have molecular structure and well studied in OBE potential model, we fit the values of the constant $B_{0}$ and $C_{0}$ to get approximate binding energy of deuteron. Then, to generalize the model to dimesonic system, we arrive on the relation between B and C in accordance of the mass of the system,   
if $ m_{m}=nm_{d}$, where $m_{m}$ is the threshold mass of dimesonic state and $m_{d}$ mass of deuteron and n is integer number (n=1,2,3..), then we could write the relation as  
\begin{eqnarray}
B &=& \frac{B_{0}}{n}  ; \ \ \ \ C = n C_{0}
\end{eqnarray}  
Here, $B_{0}$ and $C_{0}$ are the constant, fitted for deuteron binding energy. \\

\begin{figure}[h]
(a)\includegraphics[scale=0.6]{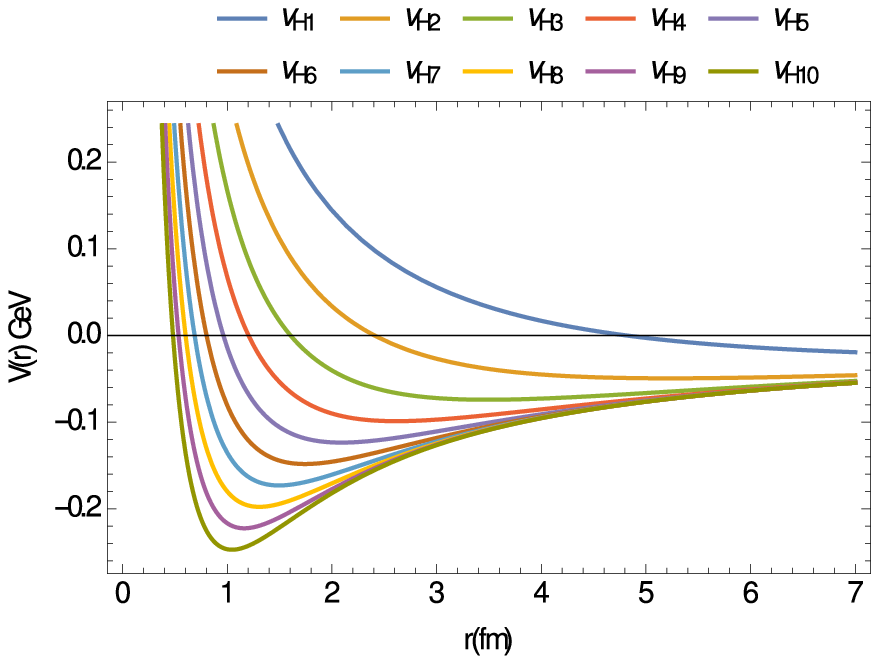} 
(b)\includegraphics[scale=0.6]{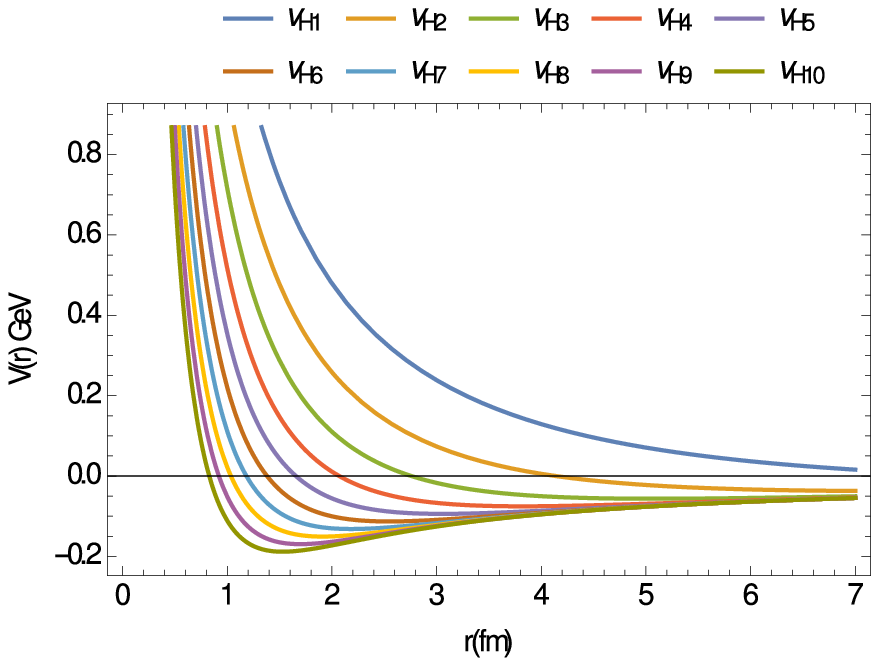} 
(c)\includegraphics[scale=0.6]{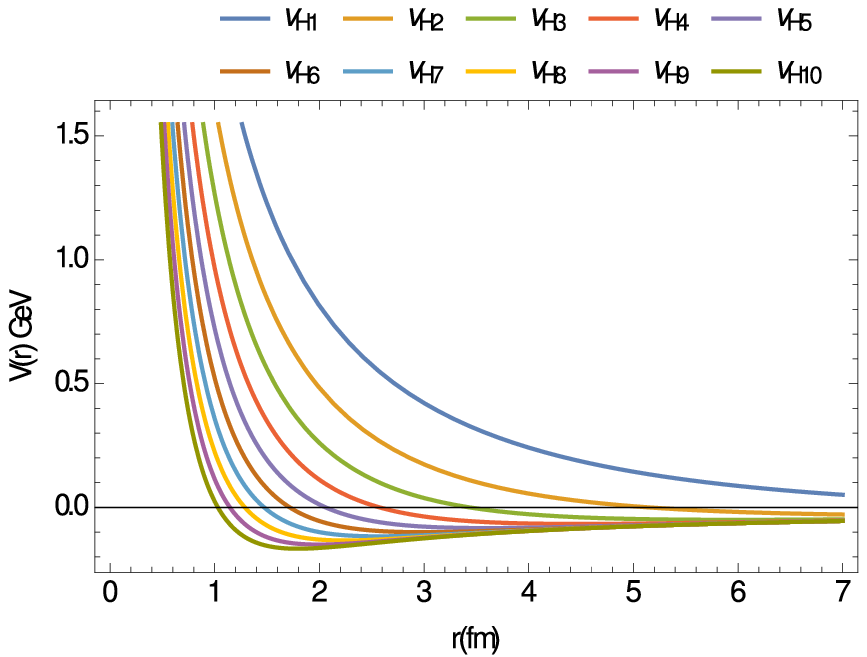} 
(d)\includegraphics[scale=0.6]{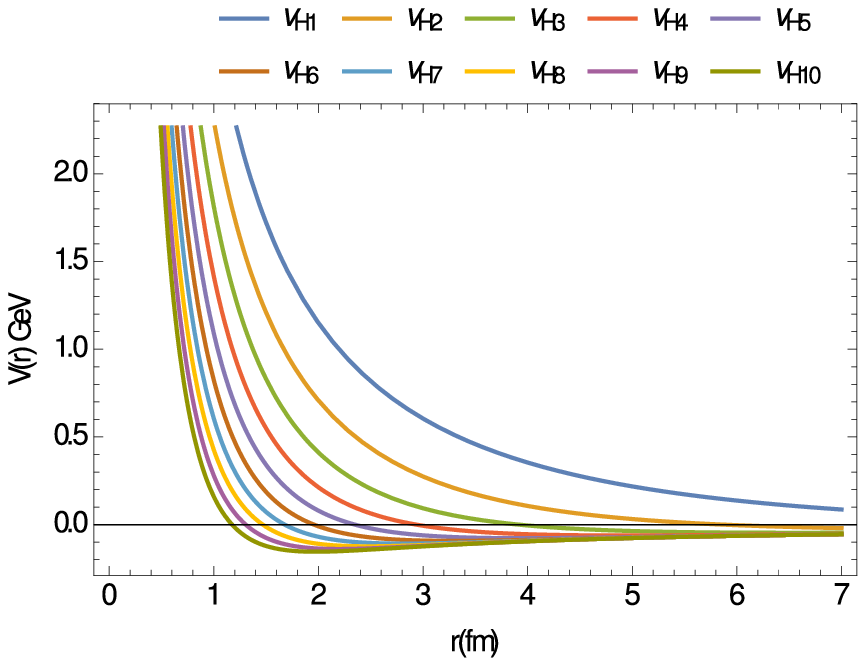} 
(e)\includegraphics[scale=0.6]{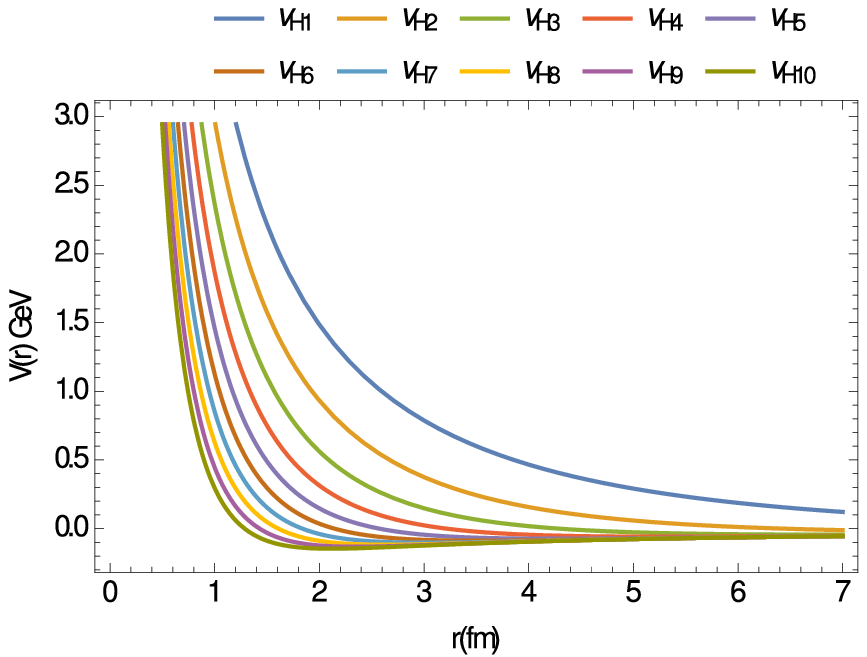}
(f)\includegraphics[scale=0.6]{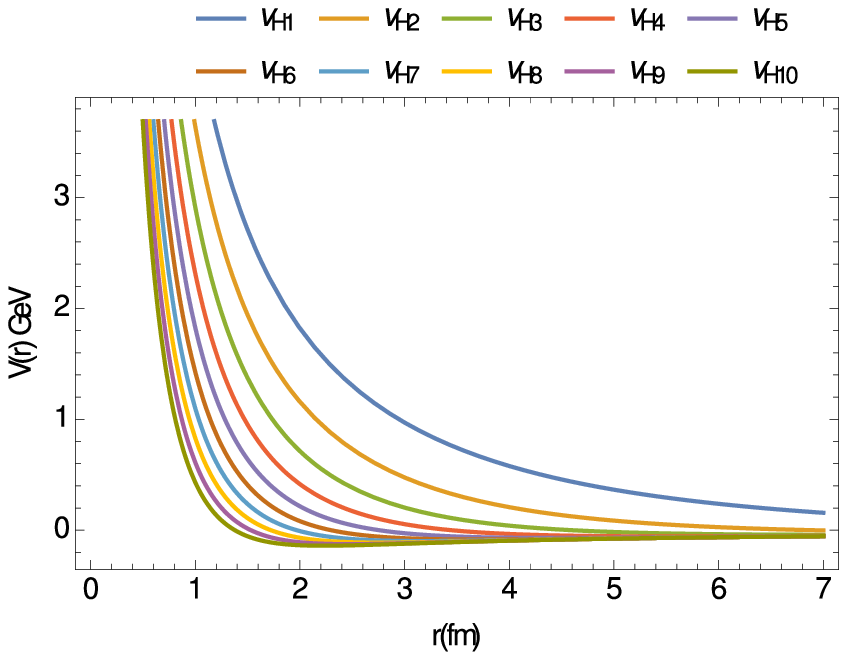} 
(g)\includegraphics[scale=0.6]{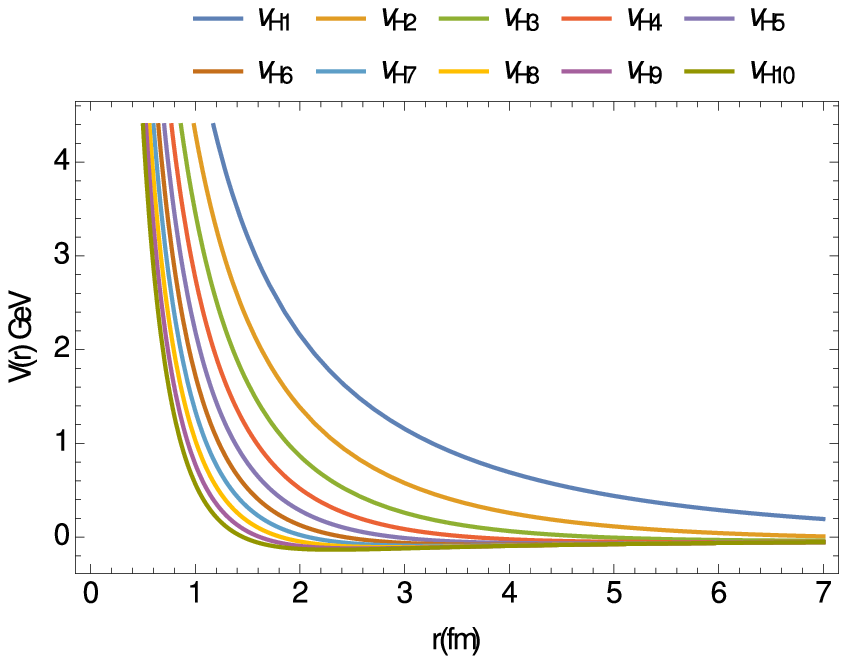} 
(h)\includegraphics[scale=0.6]{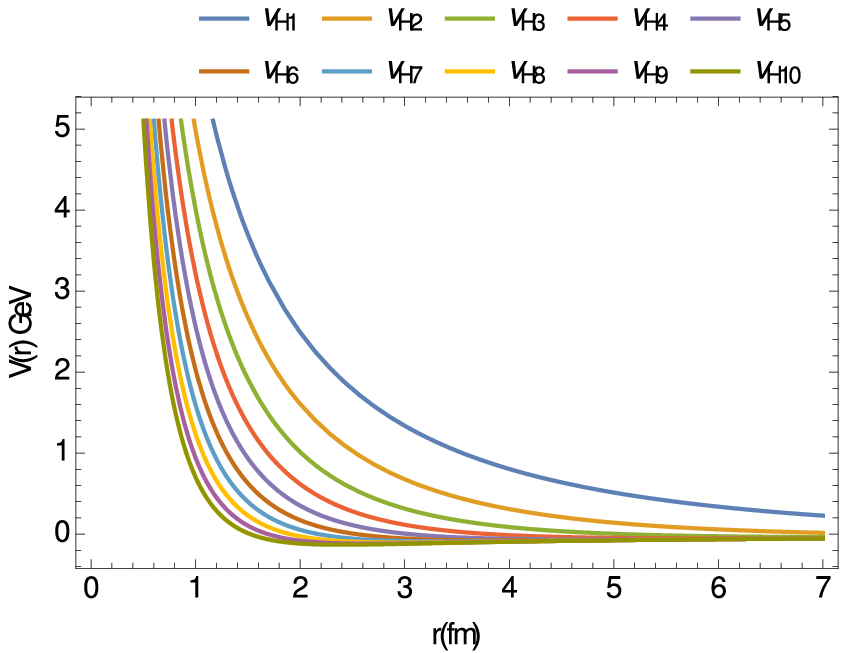} 
(i)\includegraphics[scale=0.6]{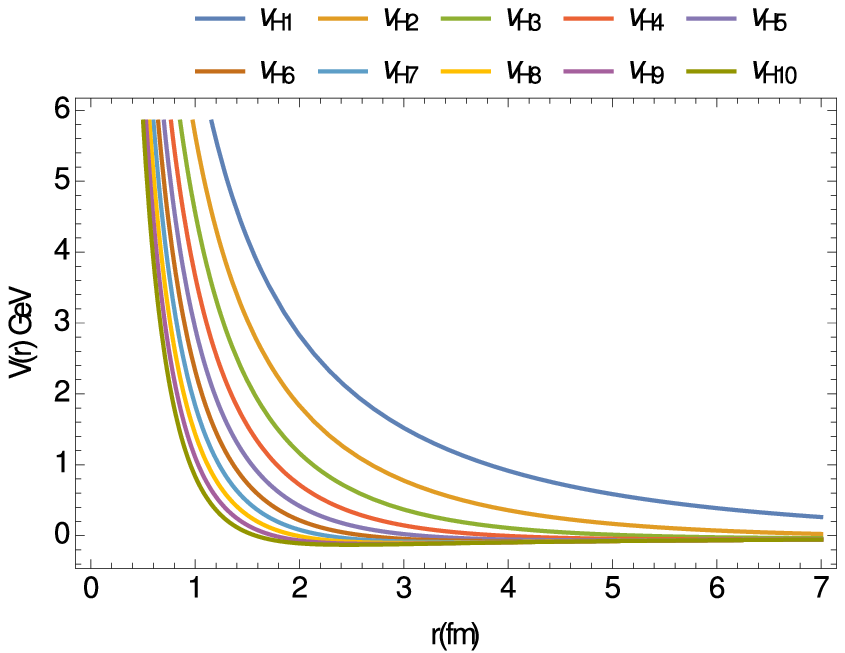} 
(j)\includegraphics[scale=0.6]{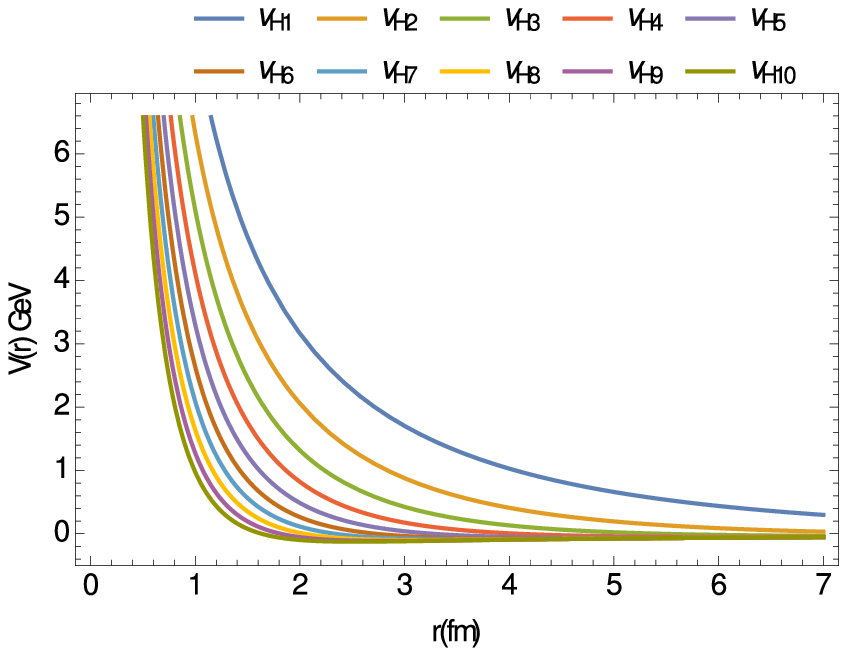} 
\caption{The figure shows the characteristic of the Hellmann potential with set of values of B and C=0.2,0.4,...2.0. (a)B=1 (b)B=2 (c)B=3 (d)B=4 (e)B=5 (f)B=6 (g)B=7 (h)B=8 (i)B=9 (j)B=10. the legends Vh1, Vh2,..Vh10 indicates the different values of C respectively, for fixed value of B. }
\end{figure}

\section{One Pion Exchange Potential}
The One Pion Exchange Potential (OPEP) taken for long range interaction which is well studied for NN-interaction. The OPE Potential for NN-interaction takes the form \cite{Paris,Pandharipande}
\begin{eqnarray}
V_{\pi}(r)&=&\frac{g_{N}^{2}}{4\pi}\frac{m_{\pi}}{3}\left(\tau_{i}\cdot\tau_{j}\right)\nonumber \\ & & 
\left[T_{\pi}(r) S_{12}+\left(Y_{\pi}(r)-\frac{4\pi}{m_{\pi}^{3}}\delta(r)\right)\left(\sigma_{i}\cdot\sigma_{j}\right)\right]
\end{eqnarray}
Where $g_{N}$ is the nucleon-pion coupling constant, $\sigma$ and $\tau$ are Spin and Isospin factors respectively while the $T_{\pi}(r)$ and $Y_{\pi}(r)$ are defined as
\begin{equation}
T_{\pi}(r)=\left(1+\frac{3}{m_{\pi}r}+\frac{3}{m_{\pi}^{2}r^{2}}\right)\frac{e^{-m_{\pi}r}}{m_{\pi}r}
\end{equation}
\begin{equation}
Y_{\pi}(r)=\frac{e^{-m_{\pi}r}}{m_{\pi}r}
\end{equation}

Whereas the $S_{12}$ is the usual tensor operator expressed as $S_{12}=3\sigma_{i}\cdot \hat{r}\sigma_{j}\cdot \hat{r}-\sigma_{i}\cdot \sigma_{j}$, is mainly responsible for long range tail of the potential and play very crucial role in the NN-interaction. The expression of OPEP in the Eq.(19) is for the point like pion. While, in a more realistic picture where the pion itself has its own internal structure, it is natural to introduced the usual form factor due to the dressing of the quarks. Hence, The form factor\cite{Pandharipande},
\begin{equation}
F^{2}_{\pi NN}(q)=\frac{\Lambda^{2}}{\Lambda^{2}+q^{2}}
\end{equation}
It is unity in the limit $q\rightarrow 0$, as required by the normalization in the above equation, and goes to zero for $q^{2}\gg \Lambda^{2}$ \cite{Pandharipande}. By getting the Fourier transform we get form factor in the r-space.  By introducing this finite size effect \cite{Paris,Pandharipande}, we have

\begin{equation}
T_{\Lambda}(r)=\left(1+\frac{3}{\Lambda r}+\frac{3}{\Lambda^{2}r^{2}}\right)\frac{e^{-\Lambda r}}{\Lambda r} \nonumber
\end{equation}
\begin{equation}
Y_{\Lambda}(r)=\frac{e^{-\Lambda r}}{\Lambda r} \nonumber
\end{equation}  
Thus, the function $T(r)$ and $Y(r)$ with the finite size effect and renormalized by strength factors $\frac{\Lambda^{2}}{\Lambda^{2}-m^{2}_{\pi}}$ and $\frac{\Lambda^{3}}{m^{3}_{\pi}}$ take the form \cite{Pandharipande}
\begin{equation}
Y(r)=\frac{\Lambda^{2}}{\Lambda^{2}-m_{\pi}^{2}}\left[Y_{\pi}(r)-\frac{\Lambda^{3}}{m_{\pi}^{3}}Y_{\Lambda}(r)\right]
\end{equation} 
\begin{equation}
T(r)=\frac{\Lambda^{2}}{\Lambda^{2}-m_{\pi}^{2}}\left[T_{\pi}(r)-\frac{\Lambda^{3}}{m_{\pi}^{3}}T_{\Lambda}(r)\right]
\end{equation}

Now, one pion exchange potential for dimesonic system could be written as
\begin{eqnarray}
V_{\pi}(c)&=& \frac{1}{12}\frac{g_{0}^{2}}{4\pi}\left(\frac{m_{\pi}}{m_{m}}\right)^{2}\left(\tau_{1}\cdot\tau_{2}\right)\left(\sigma_{1}\cdot\sigma_{2}\right)\left(\frac{\Lambda^{2}}{\Lambda^{2}-m_{\pi}^{2}}\right) \nonumber  
\\& &  
\left(\frac{e^{-m_{\pi}r_{12}}}{r_{12}}-\left(\frac{\Lambda}{m_{\pi}}\right)^{2}\frac{e^{-\Lambda r_{12}}}{r_{12}}\right)
\end{eqnarray}

\begin{eqnarray}
V_{\pi}(t)&& = \frac{1}{12}\frac{g_{0}^{2}}{4\pi}\left(\frac{m_{\pi}}{m_{m}}\right)^{2}\left(\tau_{1}\cdot\tau_{2}\right)(S_{12})\left(\frac{\Lambda^{2}}{\Lambda^{2}-m_{\pi}^{2}}\right) \nonumber  
\\& & 
\left[\left(1+\frac{3}{m_{\pi}r_{12}}+\frac{3}{m_{\pi}^{2}r_{12}^{2}}\right)\frac{e^{-m_{\pi}r_{12}}}{r_{12}} \right. \nonumber \\
&& \left.-\left(\frac{\Lambda}{m_{\pi}}\right)^{2}\left(1+\frac{3}{\Lambda r_{12}}+\frac{3}{\Lambda^{2}r_{12}^{2}}\right)\frac{e^{-\Lambda r_{12}}}{r_{12}}\right]
\end{eqnarray}

Such that, the OPEP becomes for dimesonic systems,
\begin{equation}
V_{\pi}(r_{12})=V_{\pi}(c)+V_{\pi}(t)
\end{equation}
\begin{figure*}[b]
(a)\includegraphics[scale=0.7]{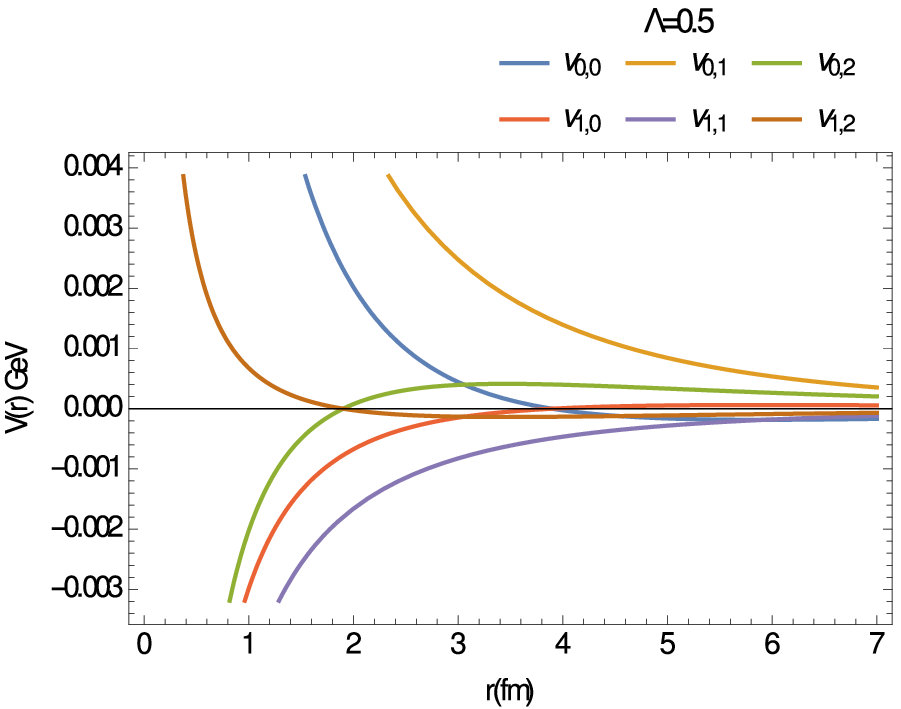} 
(b)\includegraphics[scale=0.7]{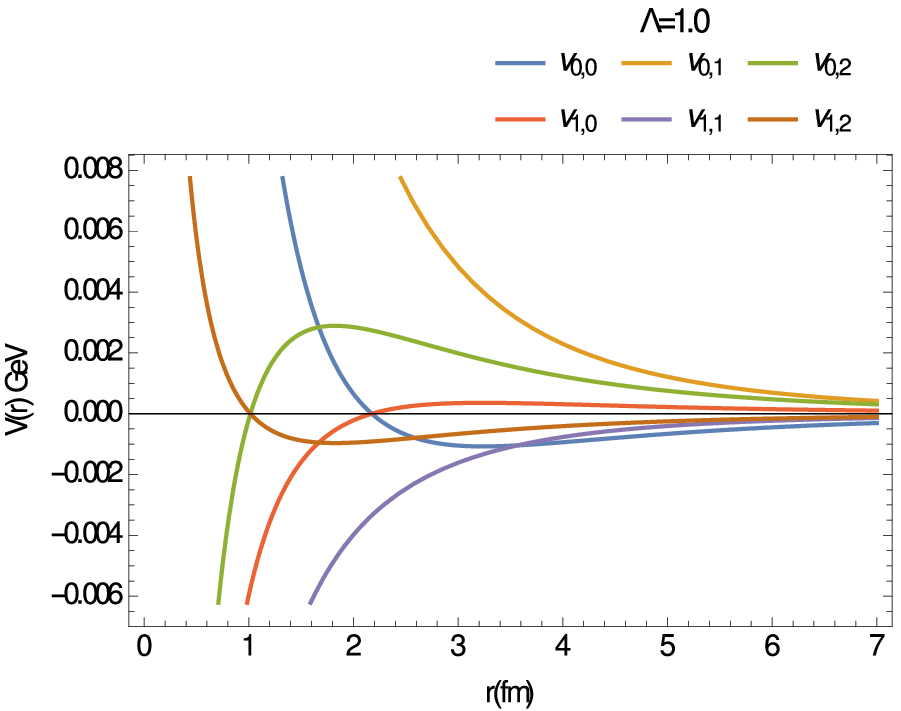}\\
(c)\includegraphics[scale=0.7]{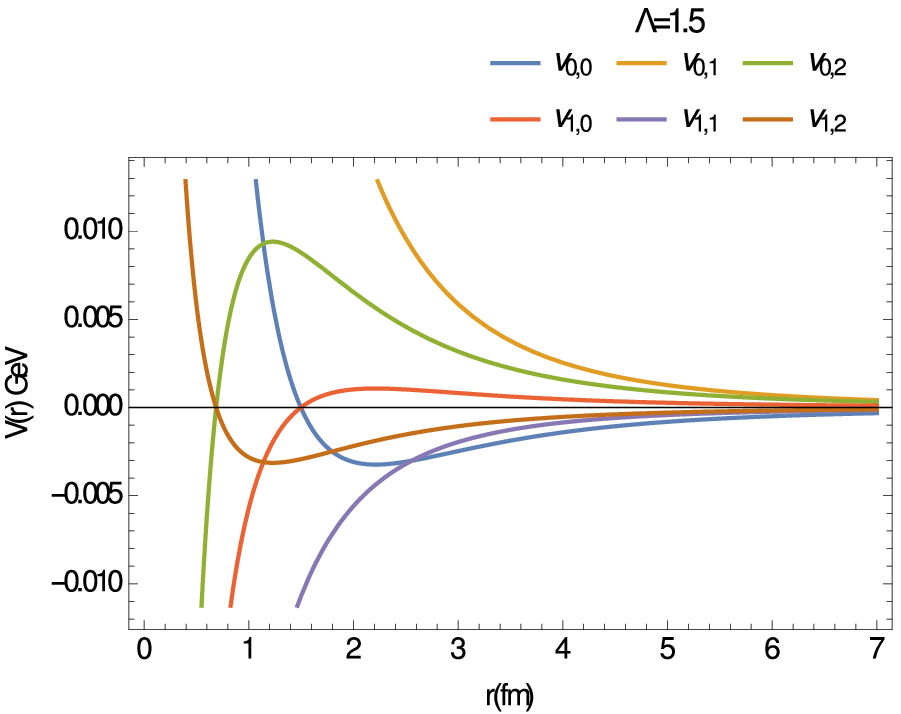}
(d)\includegraphics[scale=0.7]{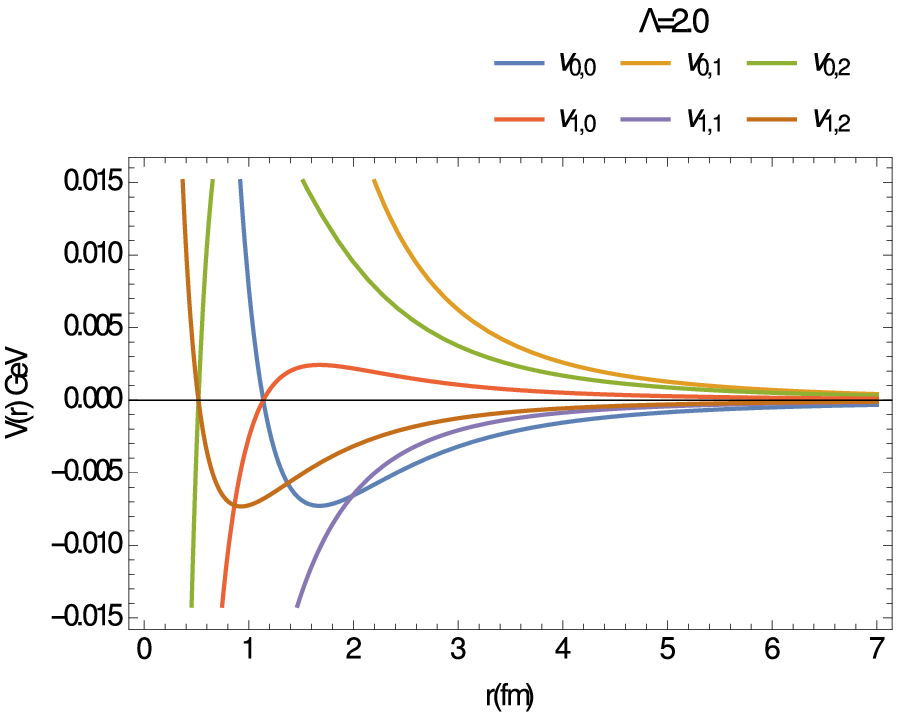}
\caption{The strength of the  One Pion Exchange Potential in different Spin-Isospin channels. The legends $V_{I,S}$ = $V_{0,0}$,$V_{0,1}$, $V_{0,2}$, $V_{1,0}$,$V_{1,1}$,$V_{1,2}$  indicates the S-wave OPEP with different vales of $\Lambda$}
\end{figure*} 

where the $g_{0}$=0.69 is the meson-pion coupling constant, $m_{m}$ and $m_{\pi}$ are the average mass of two constituent of the dimesonic state and pion mass respectively, while $\Lambda$ is the range parameter appeared in the pion form factor. The mesons and quark masses with their quantum numbers are taken from the listing of Particle Data Group\cite{Olive}. The constituent meson-quark coupling constant may derive by using the Goldberger-Treiman relation on suitable estimates of known $\pi NN$ coupling constant. The relation between quark-boson and nucleon-boson coupling could expressed as \cite{Gui-Jun Ding,D. O. Riska}
\begin{equation}
g_{\pi qq}=\frac{3 m_{q}}{5 m_{N}}g_{\pi NN} \nonumber
\end{equation}
The effective OPE potential can be split into a central and tensor term proportional to $(\sigma\cdot\sigma)$ $(\tau\cdot\tau)$  and $S_{12}$ $(\tau\cdot\tau)$ respectively. In Ref.\cite{Thomas}, Thomas et.al. mentioned and discussed sign convention and detail calculation of spin-isospin factor. Ref.\cite{Thomas} showed the inconsistent sign convention adopted by Ref.\cite{Tornqvist-Z} in the calculation of spin-isospin factor of $V\overline{V}$($2^{++}$). Then, they derive and explain overall sign for determination of $(\sigma\cdot\sigma)$ $(\tau\cdot\tau)$. We are agreed with Ref.\cite{Thomas} and adopt the same. For total spin S, total isospin state I and charge conjugation parity C, the spin-isospin factor for central term are given by \cite{Thomas}
\begin{eqnarray}
(i) &&P\overline{V} : -C(2I(I+1)-3) \nonumber \\
(ii)&&V\overline{V} : -(S(S+1)-4)(I(I+1)-\frac{3}{2})
\end{eqnarray}   

The matrix element of the tensor operator for different spin state are real numbers and it is well discussed by T$\ddot{o}$rnqvist in Ref.\cite{Tornqvist-Z}. For example, the matrix element of $S_{12}$ in the case of deuteron 

\begin{eqnarray}
&&\left\langle ^{3}S_{1} |S_{12}| ^{3}S_{1} \right\rangle=0, \left\langle ^{3}S_{1} |S_{12}| ^{3}D_{1} \right\rangle=\sqrt{8}, \nonumber \\ 
&&\left\langle ^{3}S_{1} |S_{12}| ^{3}D_{1} \right\rangle =-2 \nonumber
\end{eqnarray}

with such interaction the OPEP potential shown in Eq(29), one need to solve the coupled channel Schr$\ddot{o}$dinger equation. In the present study, we have focus only on the S-wave spectra of dimesonic state. As such, we have taken only the s-wave tensor contribution of particular spin state. The matrix element of such tensor operator of dimesonic states are given by
 
\begin{eqnarray}
&&  V_{1^{+\pm}}:\left\langle ^{3}S_{1} |S_{12}| ^{3}D_{1} \right\rangle=-\sqrt{2},  \nonumber \\ 
&&V_{0^{++}} :\left\langle ^{1}S_{0} |S_{12}| ^{5}D_{0} \right\rangle=\sqrt{\frac{1}{2}}, \nonumber \\ 
&&V_{2^{++}}\left\langle ^{5}S_{2} |S_{12}| ^{5}D_{2} \right\rangle =-\sqrt{\frac{7}{10}}  \nonumber
\end{eqnarray}

The value of the range parameter $\Lambda$ in OPEP affects the strength of the potential very drastically. The sensitivity of the $\Lambda$ have discussed by authors of Ref.\cite{Tornqvist-Z,Thomas,Gui-Jun Ding}. In the Fig(3), the nature of the OPEP in different spin-isospin state are shown for different values of $\Lambda$ (noted that only the s-wave contribution of the tensor interaction is considered). One can analyzed from the plot, as the value of the $\Lambda$ increases the strength of the potential is also increases drastically. In Ref.\cite{Tornqvist-Z}, T$\ddot{o}$rnqvist noted the value of $\Lambda$ fall in range 0.8-1.5 GeV for fit the NN scattering data where as in case of mesonic molecule, specially, the heavy meson which are very small compared to the size of the nucleon, the larger value of $\Lambda$ is expected and the large value of $\Lambda$ increases the binding energy.  Ref.\cite{Gui-Jun Ding} shows the results are very sensitive to $\Lambda$ and the binding energy no longer monotonically increases with $\Lambda$ with OPEP model. For instant, we fixed $\Lambda$=1.5 GeV consistent with NN-scattering data and increase the dominance of the Hellmann potential.\\

The factors $(\sigma_{i}\cdot\sigma_{j})$ $(\tau_{i}\cdot\tau_{j})$ and $S_{12}$ $(\tau_{i}\cdot\tau_{j})$ makes the OPEP spin and isospin state dependent. As per the literature \cite{Tornqvist-Z,Gui-Jun Ding}, the most attractive channel is (I,S)=(0,0) while the channel (I,S)=(0,2) is the most repulsive. But, as in Fig(3), we observed the channel (I,S)=(1,1) is most attractive while (I,S)=(0,1) noted as most repulsive one. The next order of the attractive channels are (I,S)=(0,0) and (I,S)=(1,2). Whereas, the channels (I,S)=(0,1),(I,S)=(0,2) and (I,S)=(1,0) expected to be unbound. The observed change in the nature of the these channels may be due to only the s-wave contribution of the tensor interaction. It clearly indicates the dominant effect of tensor operator at long range part of the potential. The discussed effect of OPEP is also reflect in our results, tabulated in Table-II and Table-III. 

\section{Decay Properties}
(i) For hadronic decay \cite{Feng-Kun Guo}:One of the constituent meson decay into two mesons: 
$ M \rightarrow e+a+d $\\  
Here, M is the mesonic molecule (M = d+b, where d and b are constituents) decaying into product mesons a,d,e in which  one of the meson is the constituent of the dimesonic state,  
For example, $DD^{*} \rightarrow \pi D \overline{D} $, here, $D^{*}\rightarrow \pi D$ or $\overline{D^{*}} \rightarrow \pi \overline{D}$ \\
From Ref \cite{Feng-Kun Guo}, On the basis of the tree level approximation the amplitude is given by  \cite{Feng-Kun Guo}
\begin{eqnarray}
{\cal{T}}_{tree}&=& -2i\frac{g g_{1}}{f_{\pi}}\sqrt{m_{m}}m_{d}m_{b}\overrightarrow{\epsilon}_{m}\dot{}\overrightarrow{P_{\pi}}\nonumber \\
&& \left(\frac{1}{P_{ae}^{2}-m_{b}^{2}}+\frac{1}{P_{de}^{2}-m_{b}^{2}}\right)
\end{eqnarray}
Where g=0.69 the quarks meson coupling constant, $f_{\pi}$= 92.2 MeV pion decay constant where as $g_{1}$=0.35 $GeV^{\frac{-1}{2}}$ is the coupling constant to the dimesonic state to the charge or neutral channels \cite{Feng-Kun Guo}. Thus, decay formula reads
\begin{eqnarray}
\Gamma_{j}&&= \frac{g^{2}}{192 \pi^{3} f_{\pi}^{2}}\left(\frac{g_{1} m_{d} m_{b}}{m_{m}}\right)^{2} \times \nonumber \\ 
&&\int_{({m_{a}+m_{\pi}})^{2}}^{({m_{m}-m_{a}})^{2}}dm_{ea}^{2} \int_{{(m_{ad}^{2})_{min}}}^{{(m_{ad}^{2})_{max}}} dm_{ad}^{2} \nonumber \\
&& \left(\frac{1}{P_{ae}^{2}-m_{b}^{2}}+\frac{1}{P_{de}^{2}-m_{b}^{2}}\right) |\overrightarrow{P_{\pi}}|^{2}
\end{eqnarray}
Where $\overrightarrow{P_{\pi}}$ is the three momentum of the pion. While $P_{ea}$ and $P_{de}$ are the the four momenta of ad and ac systems.
Since the amplitude is dependent on the invariant masses we have $m_{ea}^{2}$=$P_{ea}^{2}$ and 
$m_{ad}^{2}$=$(m_{m}^{2}+m_{\pi}^{2}+2m_{d}^{2}-m_{ea}^{2}-P_{de})$ of the final state ea and ad pairs respectively \cite{Feng-Kun Guo}. 
Here, 
\begin{eqnarray}
|\overrightarrow{P_{\pi}}|=\frac{\sqrt{\lambda(m_{m}^{2},m_{ad}^{2},m_{\pi}^{2})}}{2m_{m}}
\end{eqnarray}
 where $\lambda(x,y,z)=x^{2}+y^{2}+z^{2}-2xy-2yz-2zx$ is the k$\ddot{a}$llen function. For the known value of $m_{ea}^{2}$ the range of the $m_{ad}^{2}$ could be determined by the values of momentum  $P_{a}$ is parallel or antiparallel to $P_{d}$ as
   
\begin{eqnarray}
(m_{ad}^{2})_{max}&=& (E_{a}^{*}+E_{d}^{*})^{2}- \nonumber \\
&&\left(\sqrt{E_{a}^{*2}-m_{a}^{2}}-\sqrt{E_{d}^{*2}-m_{d}^{2}}\right)^{2} \nonumber \\
(m_{ad}^{2})_{min}&=& (E_{a}^{*}+E_{d}^{*})^{2}- \nonumber \\
&&\left(\sqrt{E_{a}^{*2}-m_{a}^{2}}+\sqrt{E_{d}^{*2}-m_{d}^{2}}\right)^{2}
\end{eqnarray}
Here, 
\begin{eqnarray}
E_{a}^{*}&=&\frac{m_{ea}^{2}-m_{e}^{2}+m_{a}^{2}}{2m_{ea}} \nonumber \\
E_{d}^{*}&=&\frac{m_{m}^{2}-m_{ea}^{2}-m_{d}^{2}}{2m_{ea}}
\end{eqnarray}
are the energies of the particle a and d respectively.\\

(ii) For radiative decay \cite{Miguel}: the constituent meson decay into photon-mesons pair

The amplitude reads as (for the$ D^{*}\overline{D^{*}}$ system)
\begin{eqnarray}
|{\cal{T}}_{k}|^{2}&=& \frac{16\pi \alpha}{3}(m_{m}m_{a}m_{d})\left(g_{2}(m_{ed})\right)^{2} \nonumber \\
&& \frac{\left(\beta_{1}+\frac{2}{3m_{c}} \right)^{2}}{\left(m_{ed}-m_{a}+i\epsilon\right)^{2}}\overrightarrow{P_{\gamma}}^{2}
\end{eqnarray}

whereas $|\overrightarrow{P_{\gamma}}|$ is given in the molecule's rest frame by
\begin{eqnarray}
|\overrightarrow{P_{\gamma}}|= E_{\gamma}=\frac{m_{m}^{2}-m_{da}^{2}}{2m_{m}}
\end{eqnarray}

\begin{eqnarray}
\beta_{1}&=&\frac{2\beta}{3}-\frac{g^{2}m_{k}}{8\pi f_{k}^{2}}-\frac{g^{2}m_{\pi}}{8\pi f_{\pi}^{2}} \nonumber \\&& and \nonumber \\
g_{2}(m_{ed})&=& g_{1} \times e^{\frac{-m_{a}(m_{a}-m_{ed})}{\Lambda_{2}}^{2}}
\end{eqnarray}
Here, $g_{2}$ is the dimesonic coupling constant, $\alpha=1/137.036$ is the fine structure constant, $m_{c}=1.5$GeV is the charm quark mass, $\beta=1/m_{q}=1/330$ $MeV^{-1}$ ($m_{q}$ is the light quark mass), $m_{\pi}$ and $m_{k}$ is the pion and kaon masses, respectively. Whereas, $f_{\pi}$ and $f_{k}$ are the pion and kaon decay constant, respectively. 
Thus the partial radiative decay width given as 

\begin{eqnarray}
\Gamma_{k}&&= \frac{1}{256 \pi^{3} m_{m}^{3}} \times \nonumber \\ &&\int_{({m_{a}+m_{d}})^{2}}^{({m_{m}})^{2}}dm_{da}^{2} \int_{{(m_{ed}^{2})_{min}}}^{{(m_{ed}^{2})_{max}}}dm_{ed}^{2} |{\cal{T}}_{k}|^{2}
\end{eqnarray}

For the known value of the $m_{da}^{2}$ the range of $m_{ed}^{2}$ could be determined by the values of momentum  $P_{d}$ is parallel or antiparallel to $P_{e}$ as
\begin{eqnarray}
(m_{ed}^{2})_{max}&=& (E_{e}^{*}+E_{d}^{*})^{2}- \nonumber \\
&&\left(\sqrt{E_{e}^{*2}}-\sqrt{E_{d}^{*2}-m_{d}^{2}}\right)^{2} \nonumber \\
(m_{ed}^{2})_{min}&=& (E_{e}^{*}+E_{d}^{*})^{2}- \nonumber \\
&&\left(\sqrt{E_{e}^{*2}}+\sqrt{E_{d}^{*2}-m_{d}^{2}}\right)^{2}
\end{eqnarray}
Here, 
\begin{eqnarray}
E_{e}^{*}&=&\frac{m_{m}^{2}-m_{da}^{2}}{2m_{da}} \nonumber \\
E_{d}^{*}&=&\frac{m_{da}^{2}+m_{d}^{2}-m_{a}^{2}}{2m_{da}}
\end{eqnarray}
$E_{e}^{*}$ and $P_{e}$ are the photon energy and momentum in the $m_{da}^{2}$ cm frame, respectively. 
Whereas, the amplitude for the $B^{*}\overline{B^{*}}$ read as

\begin{eqnarray}
|{\cal{T}}_{k}|^{2}&=& \frac{16\pi \alpha}{3}(m_{m}m_{a}m_{d})\left(g_{2}(m_{ed})\right)^{2} \nonumber \\
&& \frac{\left(\beta_{b}-\frac{1}{3m_{b}} \right)^{2}}{\left(m_{ed}-m_{a}+i\epsilon\right)^{2}}\overrightarrow{P_{\gamma}}^{2}
\end{eqnarray}

With $\beta_{b}$=$\beta_{1}$ for  $B^{*}\overline{B^{*}}$ system and could be calculated according to Eq.(B8). $m_{b}$=4.6 GeV is mass of the bottom meson. For more detailed of the calculations and formalism of decay calculation we refer to Ref.\cite{Feng-Kun Guo,Miguel}.
}

\end{document}